\documentclass[%
 preprint,
 superscriptaddress,
 amsmath,
 amssymb,
 aps,
]{revtex4-2}
\usepackage{graphicx}
\usepackage{dcolumn}
\usepackage{bm}

\usepackage{siunitx}
\usepackage{braket}
\usepackage{geometry}
\usepackage{bbm}

\usepackage[normalem]{ulem}

\begin{document}

\title{Coherent Acoustic Control of Defect Orbital States in the Strong-Driving Limit}

\author{B. A. McCullian}
 \affiliation{School of Applied and Engineering Physics, Cornell University, Ithaca, NY 14853, USA}
\author{V. Sharma}
 \affiliation{Laboratory of Atomic and Solid State Physics, Cornell University, Ithaca, NY 14853, USA}
\author{H. Y. Chen}
 \affiliation{Department of Physics, Cornell University, Ithaca, NY 14853, USA}
\author{J. C. Crossman}
 \affiliation{School of Applied and Engineering Physics, Cornell University, Ithaca, NY 14853, USA}
\author{E. J. Mueller}
 \affiliation{Laboratory of Atomic and Solid State Physics, Cornell University, Ithaca, NY 14853, USA}
\author{G. D. Fuchs}
 \email{gdf9@cornell.edu}
 \affiliation{School of Applied and Engineering Physics, Cornell University, Ithaca, NY 14853, USA}
 \affiliation{Kavli Institute at Cornell for Nanoscale Science, Ithaca, NY 14853, USA}

\date{\today}

\begin{abstract}We use a bulk acoustic wave resonator to demonstrate coherent control of the excited orbital states in a diamond nitrogen-vacancy (NV) center at cryogenic temperature. Coherent quantum control is an essential tool for understanding and mitigating decoherence. Moreover, characterizing and controlling orbital states is a central challenge for quantum networking, where optical coherence is tied to orbital coherence. We study resonant multi-phonon orbital Rabi oscillations in both the frequency and time domain, extracting the strength of the orbital-phonon interactions and the coherence of the acoustically driven orbital states. We reach the strong-driving limit, where the physics is dominated by the coupling induced by the acoustic waves. We find agreement between our measurements, quantum master equation simulations, and a Landau-Zener transition model in the strong-driving limit. Using perturbation theory, we derive an expression for the orbital Rabi frequency versus acoustic drive strength that is non-perturbative in the drive strength and agrees well with our measurements for all acoustic powers. Motivated by continuous wave spin resonance-based decoherence protection schemes, we model the orbital decoherence and find good agreement between our model and our measured few-to-several nanoseconds orbital decoherence times. We discuss the outlook for orbital decoherence protection.
\end{abstract}

\maketitle

\section{Introduction}\label{section: Introduction}

Quantum coherent control strategies can be used to both study and mitigate decoherence. This idea, along with the associated opportunities for quantum technologies, has spurred the development of high-fidelity quantum control over superconducing~\cite{Devoret_2013}, atomic~\cite{Saffman_2010}, and quantum dot~\cite{Hanson_2007} systems, among many others. Coherent control of solid-state defect spins has enabled the development of decoherence protection schemes~\cite{de_Lange_2010,Xu_2012,Golter_2014, MacQuarrie_2015,Anderson_2022} that can aid in precision sensing~\cite{Boss_2017,Barry_2020} and quantum networking~\cite{Togan_2010,Bernien_2012,Sipahigil_2012,Bernien_2013,Pfaff_2014,Wehner_2018,Awschalom_2018,Pompili_2021,Hermans_2022}. Here we use an acoustic wave resonator at cryogenic temperature to demonstrate coherent control of the excited orbital states in a diamond nitrogen-vacancy (NV) center in the strong-driving limit, and we explore orbital resonance-based decoherence mitigation.


Strain and electric fields can couple to a defect's orbital states, limiting quantum coherence while also creating opportunities for coherent control. On the negative side, electric field fluctuations from nearby charge traps provide the leading source of spectral diffusion in NV centers~\cite{Fu_2009}, which is problematic for quantum networking applications that rely on frequency-matched photon emission. On the positive side, both static electric fields~\cite{Bassett_2011, Acosta_2012} and quasi-static strains~\cite{Lee_2016} have been used to tune NV center optical transitions. Dynamical strains can also be used for quantum control: they can be combined with optical pulses to generate coherent Raman sidebands~\cite{Golter_2016,Chen_2018} and manipulate spin~\cite{Golter_2016_a}. Two recent results in other defects demonstrate the use of orbital interactions for quantum control. In the first, researchers used the strong strain-orbit interaction in silicon vacancy defects in diamond (SiV) to achieve spin control via dynamic strain~\cite{Maity_2020}. In the second, researchers used electric fields to control a combined orbital-spin ground state transition in the neutrally charged NV center~\cite{Kurokawa_2023}. Given that mitigating spectral diffusion remains an ongoing challenge for using NV centers as quantum networking nodes, and that coherent control can often be leveraged for decoherence protection, we can naturally ask: can coherent orbital control protect NV centers against spectral diffusion?

In this work we demonstrate coherent control within the negatively charged NV center excited-state orbital doublet manifold using gigahertz frequency acoustic waves. We study the resulting coherent dynamics of the associated optical transitions. We observe orbital Rabi oscillations driven by a resonant, multi-phonon mechanism and quantify the coherence of the orbital states under acoustic drive. The orbital Rabi oscillations are well-described by a strong-driving Hamiltonian, and the Rabi frequency measured in the time domain is consistent with the spectroscopic splitting of the optical transitions in the frequency domain. We numerically verify that our simple Hamiltonian model produces these same dynamics using quantum master equation simulations and we analytically describe them using a direct calculation of the orbital Rabi frequency with both a Landau-Zener transfer matrix approach as well as a perturbative approach in the Floquet picture. We characterize the coherence of the acoustically driven orbital states, finding at least a factor of two enhancement of the coherence time. Finally, we discuss orbital coherence enhancement by making an analogy with continuous-wave spin dynamical decoupling.

\section{Strain-orbital and acoustic-orbital interactions}\label{section: NV center strain-orbital interactions}

\begin{figure}
\centering
\includegraphics[scale=0.84]{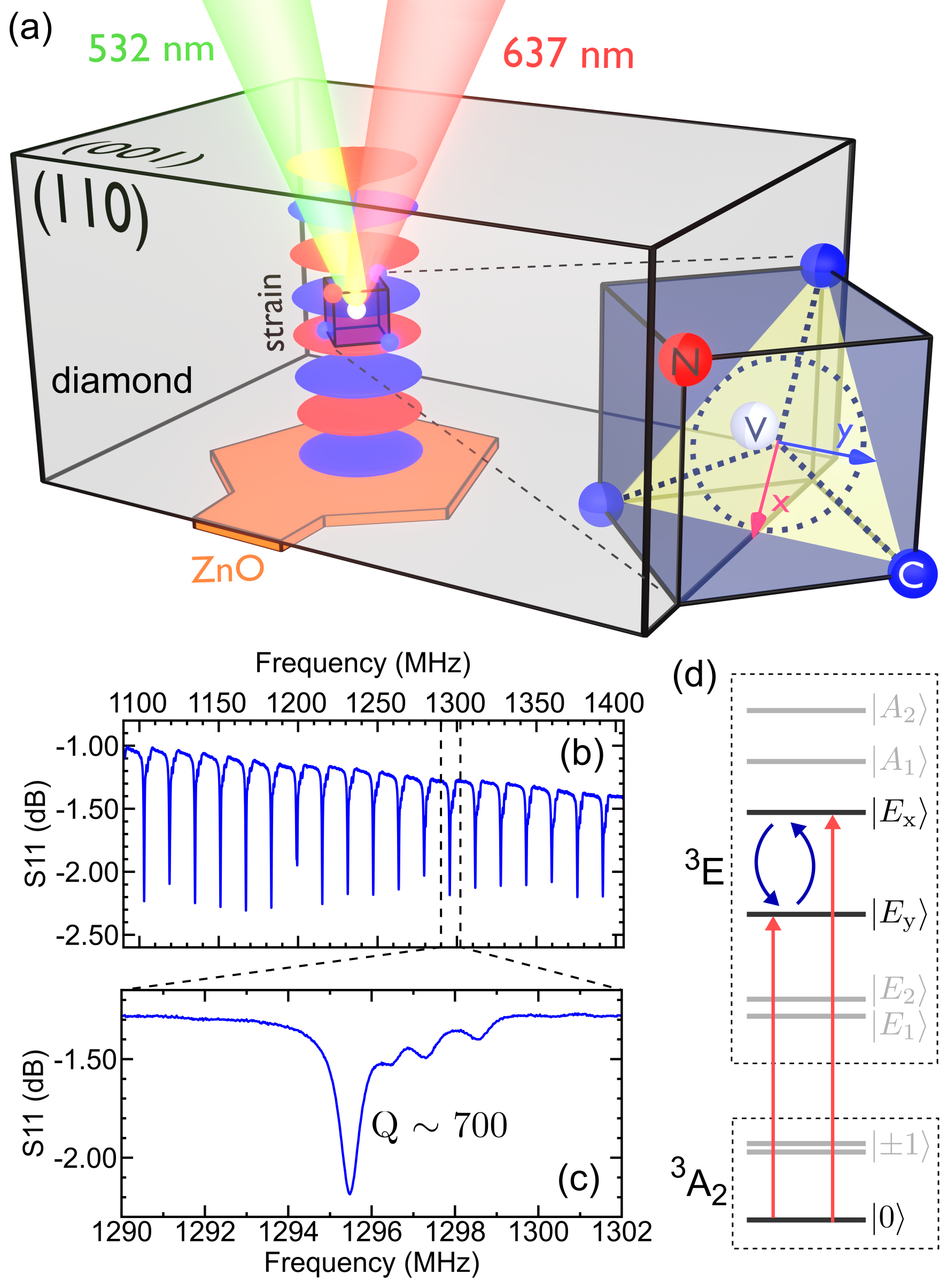}
\caption{\label{fig1} \textbf{Bulk acoustic wave resonator on diamond for acoustic orbital control.} \textbf{(a)} A ZnO transducer (orange) excites the diamond (gray) high-overtone bulk-mode acoustic resonator (HBAR) with GHz-frequency uniaxial $\sigma_{\textrm{zz}}$ strain standing waves throughout the diamond bulk (red and blue lobes). A single bulk NV center (blue unit cell) is optically excited and fluorescence is collected through the diamond surface opposite the ZnO transducer. Dipoles $x$ (red arrow) and $y$ (blue arrow) corresponding to the $\ket{0} \leftrightarrow \ket{E_{\textrm{x}}}$ and $\ket{0} \leftrightarrow \ket{E_{\textrm{y}}}$ transitions lie in a plane (yellow) normal to the NV center $C_{3,\textrm{v}}$ symmetry axis. \textbf{(b,c)} S$_{11}$ measurement of the electromechanical response of the HBAR device measured at room temperature. The quality factor of the mode used for experiments is about 700. \textbf{(d)} Energy level manifold for an NV center at low temperature. A resonant laser (red arrows) excites spin-preserving optical transitions, while acoustic driving (blue arrows) couples the orbital excited states.}
\end{figure}

We probe the acoustic driving of the excited-state orbital doublet using a tunable laser that is resonant with the spin-preserving zero-phonon optical transitions. These transitions couple the orbital singlet, spin triplet ($^3 A_2$) ground states \{$\ket{0},\ket{\pm 1}$\} to orbital doublet, spin triplet ($^3 E$) excited states \{$\ket{E_{1,2}},\ket{E_{\textrm{x,y}}}, \ket{A_{1,2}}$\} at around 1.945 eV (632.2 nm)~\cite{Davies_1976} as shown in Figure~\ref{fig1}(d). Given that the transitions do not flip the spin state, it suffices to consider the $m_{\textrm{s}}=0$ states. An unstrained NV center has $C_{3,\textrm{v}}$ point group symmetry and degenerate $\ket{0} \leftrightarrow \ket{E_{\textrm{x,y}}}$ transition energies with transition dipole orientations linking the $m_{\textrm{s}}=0$ ground and excited states as shown by the red and blue arrows in Figure~\ref{fig1}(a). Persistent static strain in the diamond lifts this degeneracy. In the basis \{$\ket{E_{\textrm{x}}}, \ket{E_{\textrm{y}}}$\} the interaction of the orbital excited states with static strain is given by:
\\
\begin{equation}\label{eqn: equation_1}
H = V_{A_1} \mathbbm{1} + V_{E_1} \sigma_{\textrm{z}} + V_{E_2} \sigma_{\textrm{x}} 
\end{equation}
\\
where $\mathbbm{1}$, $\sigma_{\textrm{z}}$, and $\sigma_{\textrm{x}}$ are the identity matrix, the $z$, and the $x$ Pauli matrices, respectively. The $V_{\lambda}$ are strain deformation potentials of Jahn-Teller~\cite{Jahn_1937} symmetry $\lambda$.

We generate the acoustic control field using a high-overtone bulk-mode acoustic resonator (HBAR) fabricated from single-crystal diamond, shown schematically in Figure~\ref{fig1}(a) (see Supplementary Information for fabrication details). Gigahertz-frequency electrical driving of the HBAR at one of the resonance modes shown in~\ref{fig1}(b,c) results in a standing longitudinal strain wave between the [001] diamond surfaces, which act as acoustic mirrors. The dynamic uniaxial strain introduces time-dependent terms into the Hamiltonian~\cite{Chen_2018}:
\\
\begin{equation}\label{eqn: equation_2}
H(t) = \big( V_{A_1} + \mathcal{A}_1 \cos (\omega_{\textrm{m}} t) \big) \mathbbm{1} + \big( V_{E_1} + \mathcal{E}_1 \cos (\omega_{\textrm{m}} t) \big) \sigma_{\textrm{z}} + V_{E_2} \sigma_{\textrm{x}}
\end{equation}
\\
where $\mathcal{A}_1$ ($\mathcal{E}_1$) is the dynamic strain driving amplitude of $A_1$ ($E_1$) symmetry and $\omega_{\textrm{m}}$ is the acoustic drive frequency. The geometry of our HBAR and NV center produces no dynamic $E_2$-symmetric potential. The acoustic drive frequency for all our experiments is $\omega_{\textrm{m}} = 2 \pi \times 1.296$ GHz.

\section{Acoustically driven Orbital States: Frequency Domain}\label{section: Frequency Domain Orbital Rabi}

\begin{figure}[t]%
\centering
\includegraphics[scale=0.84]{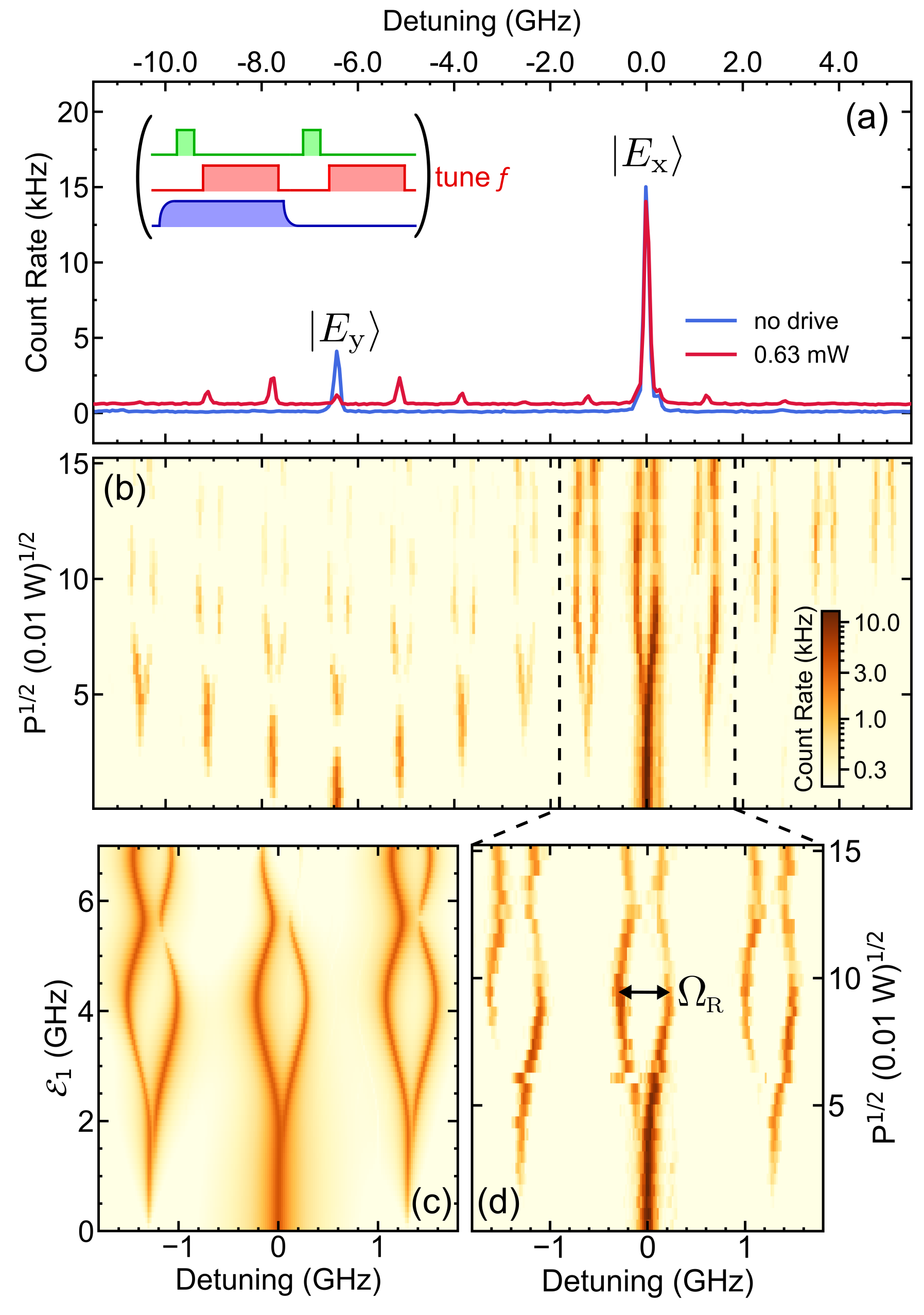}
\caption{\textbf{Acoustically driven photoluminescence excitation (PLE) spectroscopy.} \textbf{(a)} PLE spectroscopy collected with no acoustic drive (blue trace) and weak acoustic drive (red trace), showing the emergence of resolved sideband transitions. (inset) Sequence used for PLE spectroscopy measurements: 1 $\unit{\us}$ green laser, 5 $\unit{\us}$ red laser and readout, 7 $\unit{\us}$ acoustic drive, 500 ms collection at each red laser frequency. \textbf{(b)} Acoustic power dependent PLE spectroscopy with the laser frequency swept across both $\ket{E_{\textrm{x},\textrm{y}}}$ reveals sideband evolution and emergence of Autler-Townes splitting. \textbf{(c)} Simulated spectroscopy near the undriven $\ket{0} \leftrightarrow \ket{E_{\textrm{x}}}$ transition frequency is in good agreement with \textbf{(d)} zoomed in spectrum. As depicted in \textbf{(d)}, the orbital Rabi rate $\Omega_{\textrm{R}}$ can be read off as a mode splitting. The acoustic mode frequency $\omega_{\textrm{m}}$ is $2 \pi \times 1.296$ GHz and the splitting between $\ket{E_{\textrm{x}}}$ and $\ket{E_{\textrm{y}}}$ is $\Delta$ = 6.41 GHz.}\label{fig2}
\end{figure}

We begin by measuring the zero-phonon line optical transitions of a single bulk NV center using photoluminescence excitation (PLE) spectroscopy~\cite{Jelezko_2002} (see Supplementary Information for experimental details). All photoluminescence measurements are taken at zero applied magnetic field and at a temperature of 7 K. We tune a resonant laser across the two $m_{\textrm{s}}=0$ transition frequencies and collect the phonon sideband emission. We observe two peaks corresponding to the $\ket{0} \leftrightarrow \ket{E_{\textrm{x,y}}}$ transition frequencies (Figure~\ref{fig2}(a)) due to frequent re-initialization of the ground state spin of the defect to $\ket{0}$~\cite{Lee_2016}.

We then perform the same measurement in the presence of acoustic driving (0.63 mW to the transducer) and observe the emergence of coherent Raman sideband transitions at detunings of $\pm n \omega_{\textrm{m}}$ with respect to the undriven resonance frequencies ~\cite{Golter_2016,Chen_2018}. The sidebands result from $\mathcal{A}_1$ modulation, allowing for optical transitions when the energy of one laser photon $\pm$ $n$ acoustic phonons matches the undriven transition energies~\cite{Albrecht_2013}. The sidebands are resolved since $\omega_{\textrm{m}}$ exceeds the linewidth of the defect's optical transitions. We perform these measurements on several NV centers until we find one with an undriven $\ket{E_{\textrm{x}}} \leftrightarrow \ket{E_{\textrm{y}}}$ splitting that is an integer multiple of the standing wave generated by our acoustic drive.

Matching a multiple of our acoustic drive frequency to the defect's strain splitting allows us to resonantly couple the orbital states. The splitting between $\ket{E_{\textrm{x}}}$ and $\ket{E_{\textrm{y}}}$ for the defect shown in Figure~\ref{fig2} is $\Delta = 2 \sqrt{V_{E_1}^2 + V_{E_2}^2} = 6.41$ GHz and $\Delta \approx 5 \omega_{\textrm{m}}$. Our spectroscopy measurements provide a readout of the coupling strength. As we increase the acoustic drive power, we observe splitting of the PLE lines caused by resonant $n$-phonon driving of the $\ket{E_{\textrm{x}}} \leftrightarrow \ket{E_{\textrm{y}}}$ orbital transition (Figure~\ref{fig2}(b)). This Autler-Townes splitting is the spectral signature of orbital Rabi oscillation, and the line splitting gives the Rabi oscillation frequency $\Omega_{\textrm{R}}$. This oscillation originates from the $\mathcal{E}_1$ driving in the presence of a nonzero $V_{E_2}$~\cite{Chen_2018}. In a dressed state picture, it is interpreted as a resonant multi-phonon transition between the states $\ket{E_{\textrm{x}},m}$ and $\ket{E_{\textrm{y}},m+5}$ where $m$ is the number of phonons dressing the optical transition.

We characterize the strength of the acoustic drive by comparing the spectroscopic splitting with simulation. We collect a series of acoustic power dependent PLE scans in the vicinity of the undriven $\ket{E_{\textrm{x}}}$ frequency in Figure~\ref{fig2}(d). For each acoustic power we fit the peak locations of the split spectrum and extract $\Omega_{\textrm{R}}$. We find agreement between the measured spectrum and quantum master equation simulations of the acoustically driven PLE spectrum (Figure~\ref{fig2}(c)) using the QuTiP python software~\cite{Johansson_2012} (see Supplementary Information) for the Hamiltonian given in Equation~\ref{eqn: equation_2}. Thus, we can convert the applied acoustic drive power into a dynamic strain potential in units of GHz.

\section{Acoustically driven Orbital States: Time Domain}\label{section: Time Domain Orbital Rabi}

\begin{figure}[t]%
\centering
\includegraphics[scale=0.84]{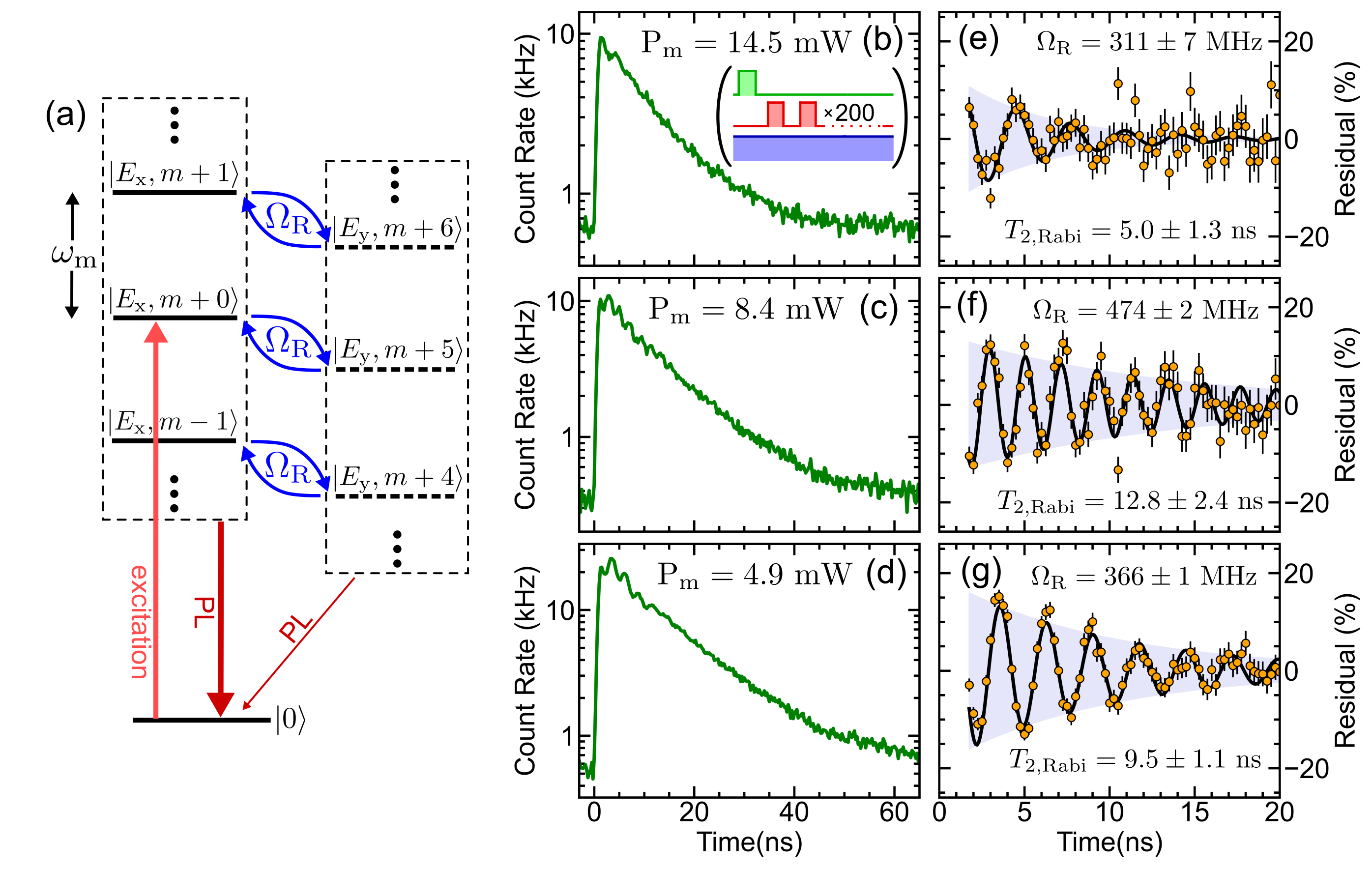}
\caption{\textbf{Time domain orbital Rabi oscillations.} \textbf{(a)} Scheme for measuring orbital Rabi oscillations in the time domain. A 1 ns duration excitation laser pulse tuned resonantly to the undriven $\ket{0} \leftrightarrow \ket{E_{\textrm{x}}}$ transition excites into the $\ket{E_{\textrm{x}}}$ orbital manifold via the $\ket{E_{\textrm{x}},m+0}$ transition. Continuous-wave acoustic driving causes orbital Rabi oscillations between the $\ket{E_{\textrm{x}}}$ and $\ket{E_{\textrm{y}}}$ orbital states. The collected PL rate from $\ket{E_{\textrm{x}}}$ is larger than the rate from $\ket{E_{\textrm{y}}}$. \textbf{(b-d)} Histogram of time-tagged photon counts relative to the repeated 1 ns duration red laser pulses for acoustic drive powers $P_{\textrm{m}}$ = 14.5 mW, 8.4 mW, and 4.9 mW. (inset) Pulse sequence: 2 $\unit{\us}$ green laser, 200 repetitions of 1 ns red laser with 100 ns of delay between red pulses. Each histogram contains 2 minutes of collected counts. \textbf{(e-g)} Extracted residual oscillations for each acoustic power. Fits to decaying sinusoid (black trace) are used to extract orbital Rabi frequency $\Omega_{\textrm{R}}$ and orbital Rabi coherence time $T_{2,\textrm{Rabi}}$. Error bars in (d-f) are determined assuming that the photon collection is shot-noise limited.}\label{fig3}
\end{figure}

We also study orbital Rabi oscillations in the time domain, which allows us to directly characterize Rabi frequencies and associated coherences. The orbital Rabi frequencies measured in Figure~\ref{fig2}(c) correspond to oscillation periods of a few nanoseconds, which is sufficiently fast that several oscillations will occur during the ~12 ns excited state lifetime~\cite{Batalov_2008,Goldman_2015}. Our scheme for measuring time-domain orbital Rabi oscillations is given in Figure~\ref{fig3}(a). We create an excited-state population primarily in the orbital state $\ket{E_{\textrm{x}}}$ by setting the laser frequency resonant to the $\ket{0} \leftrightarrow \ket{E_{\textrm{x}},m}$ transition and setting the polarization to minimize coupling into $\ket{E_{\textrm{y}},m+5}$. We resonantly excite the NV center with intense 1-ns-duration optical pulses that populate the excited state in a time that is shorter than the orbital dynamics. Given the fact that the acoustic drive is always on, longer pulses would be unable to selectively populate only one orbital state. Separate optical Rabi oscillation measurements with the laser tuned to the $\ket{0} \leftrightarrow \ket{E_{\textrm{x}}}$ transition without acoustic driving confirm that we achieve a significant excited state population with a 1 ns long optical pulse (see Supplementary Information).

Time-domain measurements of acoustically driven orbital Rabi oscillations are shown in Figure~\ref{fig3}(b-d). We record a histogram of photon arrivals relative to the onset of the excitation pulses for various acoustic drive powers. We observe a roughly exponential decay of photoluminescence as the defect undergoes spontaneous emission to the ground state. Dividing out the spontaneous emission response leaves us with an oscillatory residual, shown in Figure~\ref{fig3}(e-g) as a percentage of the PL rate which allows us to measure the orbital oscillations out to longer timescales than the spontaneous emission lifetime. We attribute these oscillations to a difference in the emission polarization of the two orbital states and their efficiency in reaching our photon detector~\cite{Kaiser_2009}. Thus, these residuals represent a direct time-domain measurement of the orbital Rabi oscillations. We fit the residuals to a decaying sinusoid of the form $y(t) = A \cos (\Omega_{\textrm{R}} t) e^{-t / T_{2,\textrm{Rabi}}}$ to determine $\Omega_{\textrm{R}}$ and the orbital Rabi coherence time $T_{2,\textrm{Rabi}}$. We observe a non-monotonic evolution of $\Omega_{\textrm{R}}$ with acoustic drive power that matches the behavior found in spectroscopy. The full set of time-domain measurements and residuals is given in the Supplementary Information.

\section{Perturbative expansion in $V_{E_2}$}\label{section: Calculating Omega}

The particular NV center in our experiment has $V_{E_2}\ll V_{E_1}$ which, as detailed in the Supplementary Information, allows us to derive a simple model for the $E_{xy}$  manifold, including an expression for $\Omega_R$ which is non-perturbative in the drive strength.  

The crux of the calculation involves transforming to a rotating frame and introducing a Floquet ansatz.  Second order perturbation theory then maps the dynamics onto an undriven two-level system,
\begin{equation}\label{twolevel}
i\partial_t \left(
\begin{array}{c}
u_0\\
v_0
\end{array}
\right)
=\left(
\begin{array}{cc}
\delta&\Omega_0\\
\Omega_0&-\delta
\end{array}
\right)
\left(
\begin{array}{c}
u_0\\
v_0
\end{array}
\right).
\end{equation}
Here $u_0$ and $v_0$ are the dominant frequency components of the Floquet wavefunction.
Up to small corrections, $|u_0|^2$ and $|v_0|^2$ correspond to the probabilities of being in the $|E_x\rangle$ and $|E_y\rangle$ state.  The parameters are
\begin{eqnarray}\label{delta}
\delta&=& V_{E_1} - \frac{n \omega_m}{2} + \sum_{s\neq 0} \frac{|\Omega_s|^2}{s \omega_m}\\
\Omega_s&=&V_{E_2} J_{s-n}\left(\frac{2{\cal E}_1}{\omega_m}\right)\label{omega}
\end{eqnarray}
where $J_s(x)$ is the Bessel function of order $s$. A special case of Equation~\ref{omega} is $\Omega_0=V_{E_2} J_{-n}(2{\cal E}_1/{\omega_m})$. At vanishing drive,
$\delta({\cal E}_1=0)= V_{E_1}-{n\omega_m}/{2} +{V_{E_2}^2}/{n\omega_m}$, and $\Omega_0({\cal E}_1=0)= 0$.  At strong drive, $\delta({\cal E}_1\to \infty)= V_{E_1}-(n\omega_m)/2$ and $\Omega_0({\cal E}_1\to \infty)=0$.  Leading corrections are discussed in the Supplementary Information.

The Rabi frequency corresponds to the the splitting between the eigen-energies of this equation, 
\begin{equation}\label{Omr}
\Omega_R=2\sqrt{\delta^2+\Omega_0^2}.
\end{equation}
One can also consider dynamics.  If we start at time $t=0$ with the defect in the $|E_x\rangle$ state, the probability of being in the $|E_y\rangle$ state at time $t$ is 
\begin{equation}\label{precession}
P_y(t)=\frac{\Omega_0^2}{\Omega_0^2+\delta^2} \sin^2{\Omega_R t/2}.
\end{equation}
This expression is similar to Rabi's formula for the time dependence of a spin driven by a resonant microwave field~\cite{Abragam_1970}. Here $\Omega_0$ and $\delta$ play the role of the driving field and the detuning, respectively.

\section{Orbital Rabi Frequency and Coherence}\label{section: Orbital Rabi Frequency and Coherence}

\begin{figure}[t]%
\centering
\includegraphics[scale=0.84]{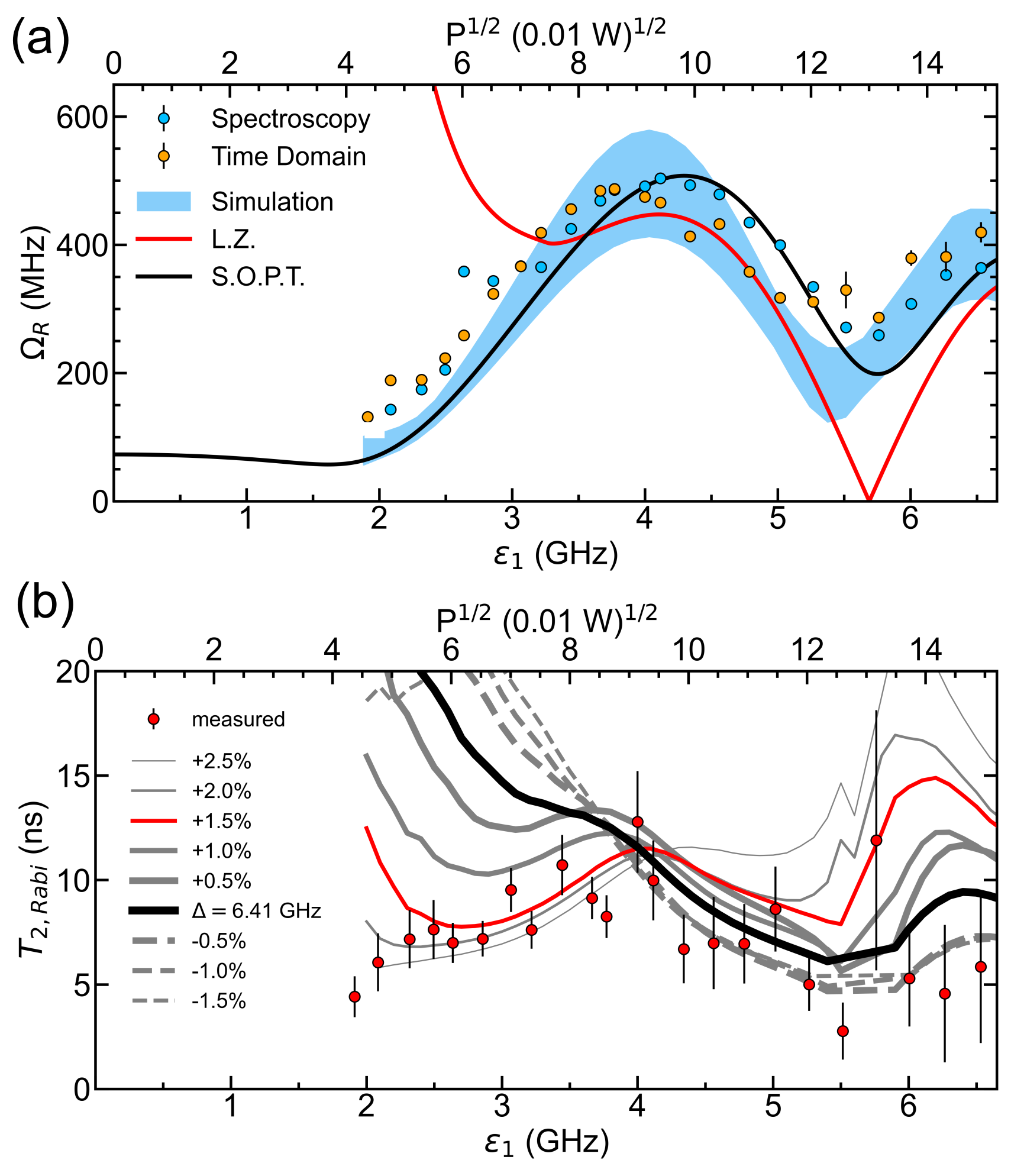}
\caption{\textbf{Orbital Rabi oscillation frequency and coherence} \textbf{(a)} $\Omega_{\textrm{R}}$ measured by spectroscopy (blue points) and in the time domain (orange points). Frequency-domain Lindblad simulation of $\Omega_{\textrm{R}}$ (blue shaded region) for $\pm$1$^\circ$ of uncertainty in defect dipole orientation and $\pm$1\% uncertainty in $\Delta$. Second order perturbation theory result for $\Omega_{\textrm{R}}$, given by Equation~\ref{Omr} (black trace), is in good agreement with the measurement for the full range of drive amplitudes. $\Omega_{\textrm{R}}$ calculated using the Landau-Zener transfer matrix method (red trace) is valid at high drive amplitudes but deviates from the experiment because it does not account for all orders of the phonon drive. \textbf{(b)} Measured $T_{2,\textrm{Rabi}}$ values (red points) extracted from the decay of acoustically driven orbital residual oscillations (Figure~\ref{fig3}(e-g)). Simulated decay time of orbital coherence for 35 MHz standard deviation, Gaussian-distributed electric field fluctuations (series of traces). The model has strong dependence on $\Delta$, and the data is best described by using a value which is 1.5\% above the nominal $\Delta$ = 6.41 GHz, which is within our experimental error in measuring $\Delta$.}\label{fig4}
\end{figure}

We compare the frequency and time domain measurements of $\Omega_{\textrm{R}}$ versus acoustic drive power. As shown in Figure~\ref{fig4}(a), they agree, confirming that our spectroscopy and time domain measurements are probing the same effect. We take several approaches to model these results. The most numerically exact of these is a quantum master equation simulation of the frequency domain response for a defect with the Hamiltonian given in Equation~\ref{eqn: equation_2} (see Supplementary Information). The blue shaded region of Figure~\ref{fig4}(a) is the simulated range of $\Omega_{\textrm{R}}$ as we vary the acoustic drive power, assuming a 1\% measurement error in the strain and optical dipole orientation of our defect. The data also agrees very well with the second order perturbation theory (SOPT) from Sec.~\ref{section: Calculating Omega} (black trace in Figure~\ref{fig4}(a)). This perturbative approach is simpler than the master equation, and gives some insight into the role being played by different components of the static strain. For example, the peaks and troughs are roughly located at the maxima and minima of $|\Omega_0|$, and the value of $\Omega_R$ at the minimum is set by $\delta$.

Hamiltonians of the same form as Equation~\ref{eqn: equation_2} have been studied in a wide variety of contexts including superconducting qubits~\cite{Oliver_2005}, quantum dots~\cite{Yang_2019}, and NV center spins~\cite{Childress_2010,Wang_2021}, with particular interest in the strong-driving limit. In this limit the amplitude of the drive $2\mathcal{E}_1$ is large compared to both the splitting $\Delta$ and the drive frequency $\omega_{\textrm{m}}$, and one can then interpret Equation~\ref{eqn: equation_2} as a periodic sequence of Landau-Zener (L.Z.) sweeps. As detailed in the Supplementary Information, this picture provides another analytic approximation to the orbital Rabi frequency (red trace in Figure~\ref{fig4}(a)). Though this approximation ignores the contributions to the acoustic driving from acoustic orders other than $n=5$, it still qualitatively agrees with the locations of the minima and maxima of $\Omega_R$ at larger ${\cal E}_1$ (red trace in Figure~\ref{fig4}(a)), confirming that we are in the strong-driving regime.

Measuring orbital Rabi oscillations in the time domain allows us to extract the orbital Rabi coherence time $T_{2,\textrm{Rabi}}$ from the exponential decay of the time domain residuals. We plot the extracted $T_{2,\textrm{Rabi}}$ versus $\Omega_{\textrm{R}}$ in Figure~\ref{fig4}(b). The orbital Rabi coherence decays on the few-to-several ns timescale, depending on the acoustic power. These timescales are similar to NV center optical Ramsey coherence times that have been previously reported~\cite{Bassett_2014}, suggesting a common origin. The coupling of our mechanical resonator to the orbital transition is sensitive to fluctuations that detune the orbital splitting from the multi-phonon acoustic drive frequency. Thus, the orbital Rabi coherence decays as a result of electric field fluctuations that are transverse to the NV center symmetry axis~\cite{McCullian_2022}. Comparatively, optical Ramsey measurements are sensitive to fluctuations that change the ground-to-excited-state transition energy which includes fluctuations both along and transverse to the NV center symmetry axis.

The decay of orbital Rabi oscillations can be understood using a model based on Equation~\ref{twolevel}. Here we assume that orbital decoherence results from fluctuations in the defect's electric field environment which modify the strain terms in Equations~\ref{delta} and \ref{omega}: $V_{E_1} \rightarrow V_{E_1} + \mathcal{E}_x$ and $V_{E_2} \rightarrow V_{E_2} + \mathcal{E}_y$. A common source of such fluctuations is shot-to-shot variation of the charge trap environment of the defect caused by the 532 nm laser, the same fluctuations which are responsible for spectral diffusion~\cite{Fu_2009}. We simulate the orbital Rabi decoherence by averaging an ensemble of sinusoidal quantum trajectories, described by Equation~\ref{precession} with a random set of electric field fluctuations $\mathcal{E}_x,\mathcal{E}_y$ drawn from a Gaussian distribution with a 35 MHz width  (see Supplementary Information).  As shown in Figure~\ref{fig4}(b), the result is quite sensitive to the splitting $\Delta$.     For ${\cal E}_1<6$ GHz we find reasonable agreement with our measurements if we increase $\Delta$ by $1.5\%$ from its nominal value of 6.41 GHz.  This deviation is within the experimental uncertainty of our spectroscopic measurement of $\Delta$. At our largest drive strengths, the decay time seems to be better fit by a model where $\Delta$ is a fraction of a percent below its nominal value.  This discrepancy could potentially be resolved by using a more sophisticated model for the electric field fluctuations, or by including further decoherence mechanisms.

\section{Conclusions and Outlook}\label{section: Conclusions and Outlook}

Previously, we have proposed the use of acoustic driving to engineer optical transition frequencies that are protected against transverse electric field noise~\cite{Chen_2018}. Additionally, our prior work indicates that many bulk NV centers are dominated by spectral diffusion sources transverse to the defect symmetry axis~\cite{McCullian_2022}. Together, these results suggest that orbital driving can mitigate spectral diffusion for some NV centers. Measuring orbital Rabi oscillations in the time domain is an essential first step toward such a decoherence protection scheme.

In the current study we find modest decoherence protection. The decoherence time is a non-monotonic function of the drive strength, and for this defect can be enhanced by at least a factor of two. The model in Section~\ref{section: Calculating Omega} lets us understand this non-monotonic behavior by making an analogy to the physics of a spin in a magnetic field under continuous microwave driving. For such a spin system, increasing the drive amplitude creates dressed states that are robust against magnetic field fluctuations, resulting in continuous-wave dynamical decoupling~\cite{Xu_2012, Golter_2014, MacQuarrie_2015}. In Equation~(\ref{twolevel}), fluctuations of $V_{E_1}$ are akin to magnetic field noise, shifting the detuning $\delta$ between the two states. Fluctuations of $V_{E_2}$ play the role of both field- and amplitude-like noise, shifting both $\delta$ and the transition matrix element, $\Omega_0$. As the acoustic drive power is varied, the decoherence is dominated by one or the other source of fluctuations. When $\Omega_R$ is large the fluctuations of $V_{E_2}$ dominate, and when $\Omega_R$ is small fluctuations of $V_{E_1}$ are most important. Thus, for this NV center, acoustic driving of the orbital states is limited to a modest improvement of the orbital coherence since both contributions to decoherence are not simultaneously negated.

Developing an orbital control scheme that mitigates all transverse fluctuations is possible by varying the device geometry. The bottleneck for our scheme is the reliance on a static $V_{E_2}$ strain to couple the orbital states (Equation~\ref{eqn: equation_2}), which results in $V_{E_2}$ fluctuations entering as amplitude-like noise in the decoherence process. Engineering a resonator-defect geometry which provides direct off-diagonal driving can result in a drive amplitude that is independent of the static strain terms, allowing for full decoherence mitigation. Additionally, electric field control of the orbital states~\cite{Kurokawa_2023} can be leveraged for pulsed orbital driving. Overall, coherent orbital control will enable researchers to apply the robust toolbox of spin resonance protocols to the orbital states and develop decoherence protection schemes that improve the single photon properties of defects.

\section*{Acknowledgements}

This work was supported by the Office of Naval Research under Grant No. N00014-21-1-2614, and by the National Science Foundation under Grant No. PHY-2110250.  Device fabrication was performed in part at the Cornell Nanoscale Facility, a member of the National Nanotechnology Coordinated Infrastructure (NNCI), which is supported by the NSF (NNCI-2025233), and at the Cornell Center for Materials Research Shared Facilities that are supported through the NSF MRSEC program (Grant No. DMR-1719875).

\clearpage

\bibliography{bibliography}

\end{document}


\title{Supplementary Information: Coherent Acoustic Control of Defect Orbital States in the Strong-Driving Limit}

\author{B. A. McCullian}
 \email{bam327@cornell.edu}
 \affiliation{School of Applied and Engineering Physics, Cornell University, Ithaca, NY 14853, USA}
\author{V. Sharma}
 \affiliation{Laboratory of Atomic and Solid State Physics, Cornell University, Ithaca, NY 14853, USA}
\author{H. Y. Chen}
 \affiliation{Department of Physics, Cornell University, Ithaca, NY 14853, USA}
\author{J. C. Crossman}
 \affiliation{School of Applied and Engineering Physics, Cornell University, Ithaca, NY 14853, USA}
\author{E. J. Mueller}
 \affiliation{Laboratory of Atomic and Solid State Physics, Cornell University, Ithaca, NY 14853, USA}
\author{G. D. Fuchs}
 \email{gdf9@cornell.edu}
 \affiliation{School of Applied and Engineering Physics, Cornell University, Ithaca, NY 14853, USA}
 \affiliation{Kavli Institute at Cornell for Nanoscale Science, Ithaca, NY 14853, USA}

\date{\today}

\maketitle

\renewcommand{\theequation}{S\arabic{equation}}
\renewcommand{\thefigure}{S\arabic{figure}}
\renewcommand{\thesection}{S\arabic{section}}

\section{Sample Fabrication}\label{section: Sample Fabrication}

The HBAR device used in our experiment was fabricated using the same procedure as our previous study~\cite{Chen_2018}. We begin with a 500 \textmu m thick type IIa diamond substrate purchased from Element Six. Isolated bulk NV centers are formed via electron irradiation and subsequent annealing. We fabricate several HBARs on the same diamond chip by sputtering a bottom metal layer [Ti(15 nm)/Pt(90 nm)], a piezoelectric layer [ZnO (1.20 \textmu m)], and a top metal layer [Ti(15 nm)/Pt(180 nm)]. The top metal layer is patterned using lift-off into several apodized pentagons that form our top HBAR electrodes. A final ZnO etch reveals the bottom metal electrode for wirebonding.

\section{PLE measurement details}\label{section: PLE measurement details}

All optical measurements are taken at zero applied magnetic field and at 7~K in a helium flow cryostat. We use confocal microscopy to address an isolated single bulk NV center through the diamond surface opposite the HBAR.

For the PLE spectroscopy measurements shown in Figure~2 of the main text, we use the same experimental sequence as in our previous study~\cite{Chen_2018}. We drive the HBAR for 7 \textmu s, allowing for 2 \textmu s of ringing up. During this ringing up time, we apply an off-resonant 1 \textmu s 532 nm laser pulse that initializes the NV center spin to $\ket{m_{\textrm{s}}=0}$ and the charge to NV$^-$. We then excite the defect with a 637.2 nm tunable laser and collect phonon sideband emission for 5 \textmu s. We also perform the same optical excitation and collection when the HBAR is not driven as a control.

For the time-resolved PLE measurements in Figure~3 of the main text, we set the red laser frequency to be resonant with the $\ket{0} \leftrightarrow \ket{E_{\textrm{x}}}$ transition. While continously driving the HBAR, we apply a sequence consisting of a 2 \textmu s green laser to initialize the defect spin and charge state, followed by 200 repetitions of a 1 ns red laser pulse. Noting that the excited state lifetime is 12 ns, we include a 100 ns delay between red laser pulses to allow the defect sufficient time to reach the ground state. We perform time correlated photon counting on the sideband emitted photons incident on our detector relative to the rising edges of the red laser pulses and record a histogram of the first photon arrival times. Each histogram in Figure~3 of the main text contains 2 minutes of collected counts (about 10$^9$ red pulses).

\section{Quantum Master Equation Simulations}\label{section: Quantum Master Equation Simulations}

We perform quantum master equation simulations of the acoustic orbital driving. These simulations help us to get an accurate conversion between the incident electrical energy on our HBAR and the amplitude of the dynamic strain term driving the orbital transitions, as well as simulate the time and frequency domain dynamics. To do so we use the master equation solver of the QuTiP package in python~\cite{Johansson_2012}.

We work in the basis of states $\{ \ket{0}, \ket{E_{\textrm{x}}}, \ket{E_{\textrm{y}}} \}$, beginning with a density matrix $\rho (t = 0)$ with all weight in the initial ground state $\ket{0}$. We time evolve the system using a quantum master equation of the form:

\begin{equation}\label{equation: sup1}
\dot{\rho}(t) = \frac{-i}{\hbar} [H(t) , \rho(t)] + \sum\limits_{n} \frac{1}{2} [ 2 C_n \rho(t) C_n^{\dagger} - \rho(t) C_n^{\dagger} C_n - C_n^{\dagger} C_n \rho(t) ]
\end{equation}
\
where $H(t)$ is the time-dependent Hamiltonian and $C_n$ are the collapse operators.

The Hamiltonian of our system is given by \cite{Chen_2018}:
\begin{equation}\label{equation: sup2}
\resizebox{.9\hsize}{!}{%
$H(t) = 2\pi
\begin{pmatrix}
\Delta_{\textrm{x}} & \Omega_{\textrm{l,x}}/2 & \Omega_{\textrm{l,y}}/2\\
\Omega_{\textrm{l,x}}/2 & V_{A_1} + V_{E_1} + (\mathcal{A}_1 + \mathcal{E}_1) \cos (\omega_{\textrm{m}} t)  & V_{E_2} \\
\Omega_{\textrm{l,y}}/2 & V_{E_2} & V_{A_1} - V_{E_1} + (\mathcal{A}_1 - \mathcal{E}_1) \cos (\omega_{\textrm{m}} t)    
\end{pmatrix}
$}
\end{equation}
\
where $\Delta_{\textrm{x}}$ is the laser detuning, $\Omega_{\textrm{l,x}}$ and $\Omega_{\textrm{l,y}}$ are the strengths of the laser coupling the ground state $\ket{0}$ to the excited states $\ket{E_{\textrm{x}}}$ and $\ket{E_{\textrm{y}}}$, respectively. $V_{i}$ are the static strain potentials of $i$ Jahn-Teller symmetry that result from the local strain environment of the defect. Going forward we will ignore $V_{A_1}$ since this term provides only a global shift of the optical transition frequencies that does not contribute to our analysis. $\mathcal{A}_1$ and $\mathcal{E}_1$ are the dynamic strain amplitudes driven by the HBAR. $\omega_{\mathrm{m}}$ = $2 \pi \ \times $ 1.296 GHz is the mechanical drive frequency.

The collapse operator matrix describing the relaxation of the defect is given by:

\begin{equation}\label{equation: sup3}
C =
\begin{pmatrix}
0 & \sqrt{\Gamma} & \sqrt{\Gamma} \\
0 & \sqrt{\gamma} & 0 \\
0 & 0 & \sqrt{\gamma}    
\end{pmatrix}
\end{equation}
\
where $\Gamma$ = 1/(12 ns) is the spontaneous relaxation rate of the NV center excited states \cite{Batalov_2008,Goldman_2015} and $\gamma$ = 1/(10 ns) sets the coherence time in the excited state orbital manifold. We choose the value of 10 ns for the coherence time based on a previous measurement of the coherence between the ground state and orbital excited states~\cite{Bassett_2014}.

Several parameters in $H(t)$ need to be quantified to carry out accurate quantum master equation simulations, namely, the static strain potentials $V_{E_1}$ and $V_{E_2}$, the mechanical drive frequency $\omega_{\textrm{m}}$, the dynamic drive amplitudes $\mathcal{A}_1$ and $\mathcal{E}_1$, the laser rates $\Omega_{\mathrm{l,x}}$ and $\Omega_{\mathrm{l,y}}$, the optical decay rate $\Gamma_{\textrm{opt}}$, and the orbital coherence rate $\Gamma_{\textrm{orb}}$. Additionally, we introduce a realistic time domain laser profile, and a simple model of spectral diffusion.

\begin{figure}
\centering
\includegraphics[scale=0.84]{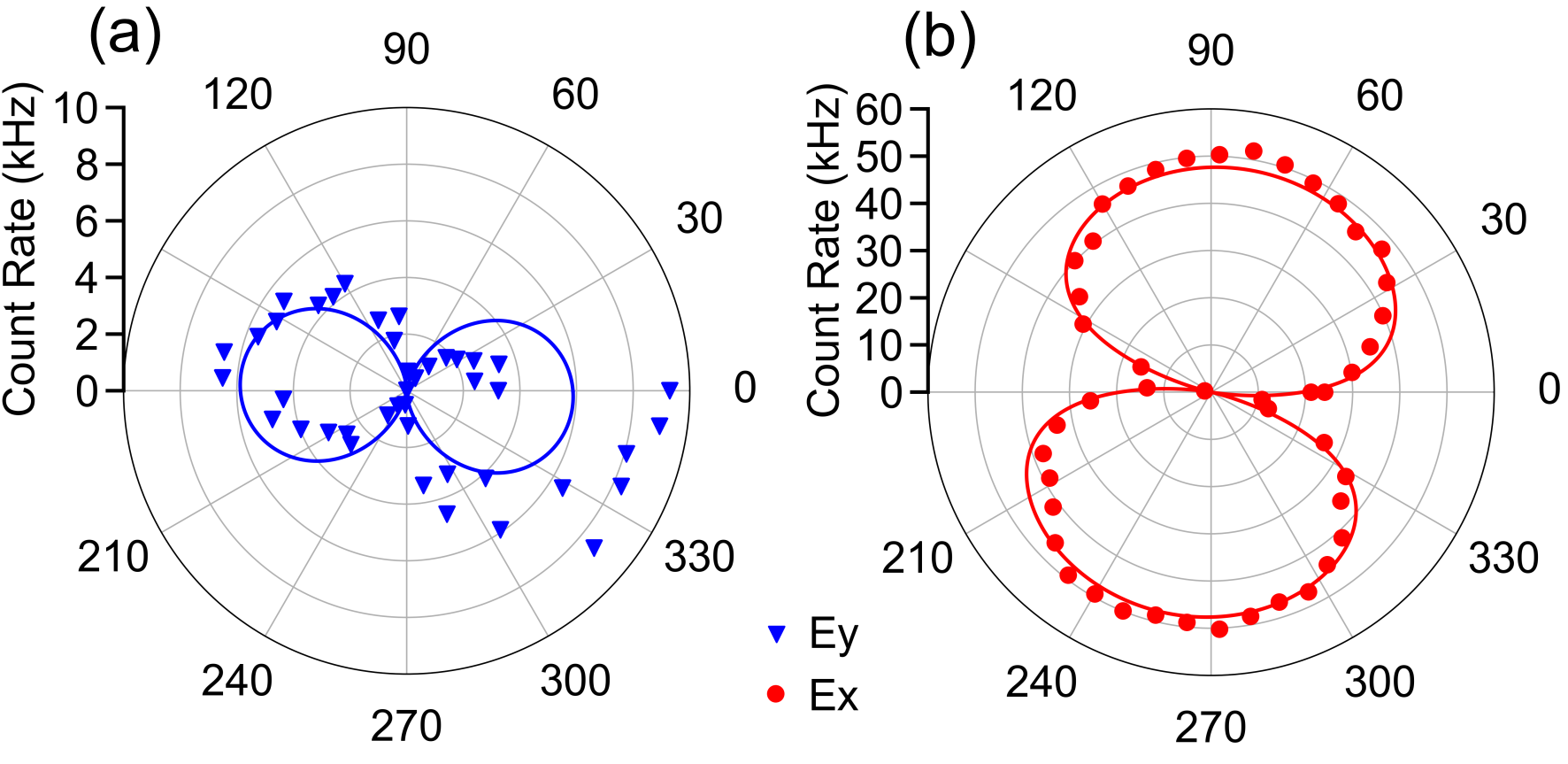}
\caption{\textbf{NV center dipole orientation measurements.} PLE count rate versus laser polarization direction for the $\ket{0} \leftrightarrow \ket{E_{\textrm{y}}}$ (blue points) and $\ket{0}$ $\leftrightarrow$ $\ket{E_{\mathrm{x}}}$ (red points) optical transitions. We extract the defect dipole orientation $\theta$ = -6.5$^\circ$ using fits to the known angular dependence (solid lines).}\label{fig:Figure_S1_version_1}
\end{figure}

We determine $V_{E_1}$ and $V_{E_2}$ by measuring the NV center optical dipole orientation as shown in Figure~\ref{fig:Figure_S1_version_1}. We tune the laser into resonant excitation with the $\ket{0} \leftrightarrow \ket{E_{\textrm{y}}}$ transition and rotate the polarization of the laser while monitoring sideband emission. We also repeat this measurement for $\ket{0} \leftrightarrow \ket{E_{\textrm{x}}}$. Using the fitting equations given in our previous study~\cite{McCullian_2022} we extract a dipole orientation $\theta$ = $-6.5^\circ$. Using the known relation between the $\ket{E_{\textrm{y}}} \leftrightarrow \ket{E_{\textrm{x}}}$ splitting, $\Delta = 2 \sqrt{V^2_{E_1} + V^2_{E_2}}$ = 6.41 GHz, and the dipole orientation, $\tan (2 \theta) = V_{E_2}/V_{E_1}$~\cite{Lee_2016}, we extract $V_{E_1}$ = $-$3.13 GHz and $V_{E_2}$ = 0.72 GHz.

\subsection{Spectroscopy Simulations}\label{subsection: Spectroscopy Simulations}

\begin{figure}
\centering
\includegraphics[scale=0.84]{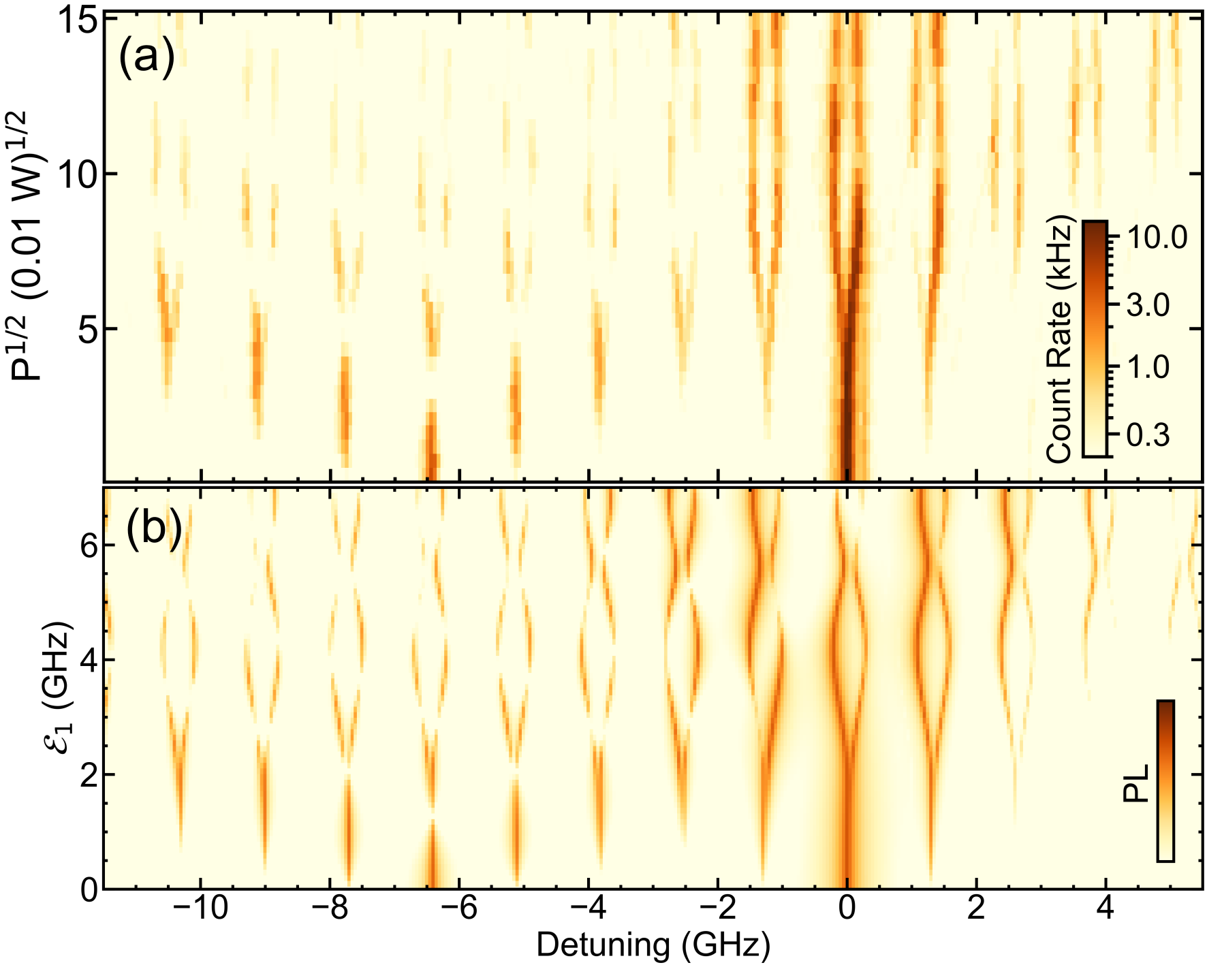}
\caption{\textbf{Determining the strength of $\mathcal{A}_1$ and $\mathcal{E}_1$.} \textbf{(a)} Measured PLE spectroscopy showing the sideband transitions evolution with increasing mechanical drive power. \textbf{(b)} Simulated PLE spectroscopy showing a close match to the sideband transitions for $\mathcal{E}_1 = (-0.7) \mathcal{A}_1$ and $\beta = 0.7$.}\label{fig:Figure_S2_version_1}
\end{figure}

Next, we determine the ratio of $\mathcal{A}_1$ to $\mathcal{E}_1$ and the conversion between applied mechanical power $P_\textrm{m}$ to $\mathcal{E}_1$ by comparing our measured PLE spectra to simulated spectra as shown in Figure \ref{fig:Figure_S2_version_1}. For this simulation we use the experimentally determined polarization of the laser to set $\Omega_{\textrm{l,x}} = \Omega_{\textrm{l,y}}$ and we set the strength of the laser field to be sufficiently weak ($\Omega_{\textrm{l}}$ = 0.05 GHz) such that we avoid significantly broadening the transitions. We calculate the density matrix over a time evolution of 50 ns and record the average values of $\rho_{11}$ and $\rho_{22}$, the $\ket{E_{\textrm{x}}}$ and $\ket{E_{\textrm{y}}}$ populations. The detected photoluminescence in our experiment is proportional to the excited state populations via the relation $\textrm{PL} = \alpha ( \rho_{11} + \beta \rho_{22} )$ where $\alpha$ is an overall scaling factor and $\beta$ is a constant which determines the ratio of detected photoluminescence counts from the two excited states~\cite{Kaiser_2009}. We adjust the ratio and range of $\mathcal{A}_1$ to $\mathcal{E}_1$ by hand until we arrive at a good match between the PLE fringes in experiment and simulation. From this, we determine that $\mathcal{E}_1 = (-0.7) \mathcal{A}_1$ and that we reach $\mathcal{E}_1$ of about 7 GHz in our experiment.

We note that because of a possible $\pm 3^\circ$ miscut of our diamond there is some uncertainty in our measurement of $\theta$. For the simulation shown in Figure 4 of the main text we use a value of $\theta$ = $-5.8^\circ$ which we hand-tune to best match the maximal orbital Rabi frequency found in experiment (Figure~2 of the main text).

\subsection{Time-Domain Simulations}\label{subsection: Time-Domain Simulations}

We now turn to time domain simulations to better understand the measured residual signals which are the result of orbital Rabi oscillations (Figure~3 of the main text).

Several experimental factors influence our time domain measurements. First, the electro-optic modulator (EOM) used for generating nanosecond-scale resonant laser pulses for our time-domain experiments has imperfect extinction. We measure the power suppression to be 21.5 dB which corresponds to a laser electric field amplitude of about 8\% of the open value when the EOM is closed. We employ a laser profile which matches our measured EOM rise time (0.75 ns) by defining the time dependent laser pulse to linearly rise and fall for this duration. We account for non-negligible laser amplitude when the EOM is gated off by applying 8\% of the laser drive strength throughout the simulation, and by performing simulations across two laser pulses separated by 100 ns, same as experiment. The first pulse takes the system from an initial pure ground state and leaves it in some mixed state determined by the strength of the imperfect laser extinction and the relaxation times. We then use the response of the density matrix to the second laser pulse for our simulation, since this best matches our experimental conditions.

Spectral diffusion also plays a key role in the photo-physics of the NV center. To account for the effects of spectral diffusion we introduce a randomly generated electric field perturbation to the Hamiltonian which is given by:

\begin{equation}\label{equation: sup4}
\begin{split}
H_j(t) & = H_{\textrm{static}} + H_{\textrm{dynamic}}(t)\\
& = 2\pi
\begin{pmatrix}
\Delta_{\textrm{x}} & 0 & 0\\
0 & E_{A_1,j} + V_{E_1} + E_{E_1,j} & V_{E_2} + E_{E_2,j} \\
0 & V_{E_2} + E_{E_2,j} & E_{A_1,j} - V_{E_1} + E_{E_1,j}  
\end{pmatrix}
\\
& + 2\pi
\begin{pmatrix}
0 & \Omega_{\textrm{l,x}}(t)/2 & \Omega_{\textrm{l,y}}(t)/2\\
\Omega_{\textrm{l,x}}(t)/2 & (\mathcal{A}_1 + \mathcal{E}_1) \cos (\omega_{\textrm{m}} t) & 0\\
\Omega_{\textrm{l,y}}(t)/2 & 0 & (\mathcal{A}_1 - \mathcal{E}_1) \cos (\omega_{\textrm{m}} t)
\end{pmatrix}
\end{split}
\end{equation}
where $E_{i,j}$ are $i$-symmetric electric field perturbations which are randomly drawn from Gaussian probability distributions with 30 MHz standard deviation. The width of this range of spectral diffusion process is chosen such that the resulting frequency-domain linewidth of the NV center is approximately 100 MHz, similar to what we measure in spectroscopy. We average over 50 randomly generated sets of $E_{i,j}$ to account for spectral diffusion. We also introduce a random phase to the mechanical drive terms for each iteration in order to remove phase-dependent simulation artifacts.

\begin{figure}
\centering
\includegraphics[scale=0.84]{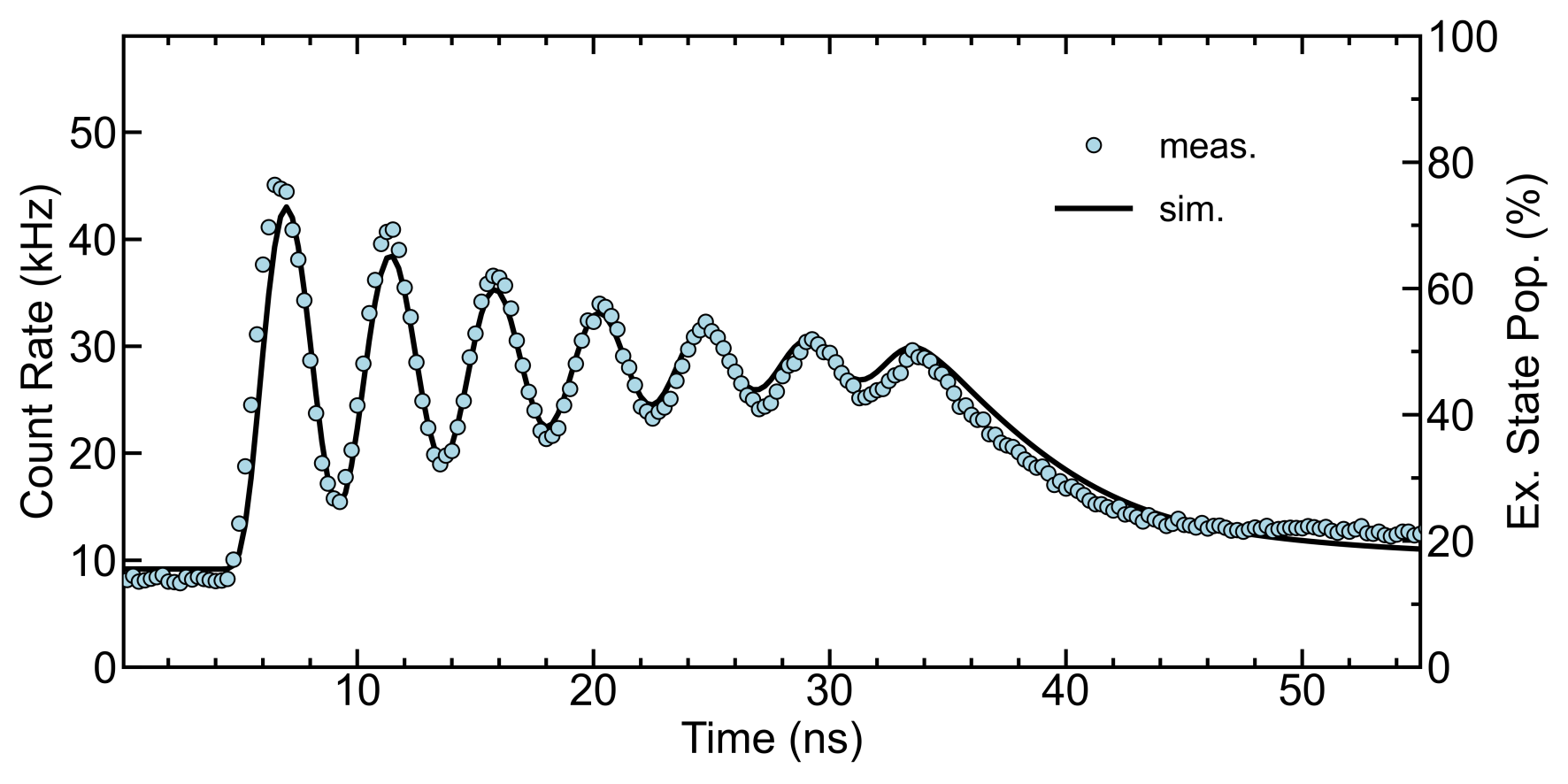}
\caption{\textbf{Determining the laser rate for simulation.} Measured histogram of time-tagged photon counts relative to a 30 ns duration red laser pulse at $\Delta$ = 0 (light blue points) shows optical Rabi between the $\ket{0}$ and $\ket{E_{\textrm{x}}}$ states. Simulated optical Rabi (black trace) for imperfect laser extinction and averaged over spectral diffusion is used to determine the optical excitation rate appropriate for time-domain simulations.}\label{fig:Figure_S3_version_1}
\end{figure}

Having dealt with the imperfect laser extinction and spectral diffusion, we next turn to determining the laser drive strength. We do so by comparing an optical Rabi measurement of the $\ket{0} \leftrightarrow \ket{E_{\textrm{x}}}$ transition with simulation, as shown in Figure~\ref{fig:Figure_S3_version_1}. For all time domain measurements we set the laser polarization to be along $\hat{x}$, and we measure the laser extinction to be 100:1 in power (and so 10:1 in electric field amplitude). Thus, for simulation we constrain $\Omega_{\textrm{l,y}} = (0.1) \Omega_{\textrm{l,x}}$ and vary only the amplitude of the laser field. We find excellent agreement between the simulated and measured optical Rabi responses when $\Omega_{\textrm{l,x}}$ = 0.22 GHz. Imperfect laser extinction from our EOM results in the count rate being non-negligible before the EOM opens at 5 ns. This effect is well captured by our simulation.

\begin{figure}
\centering
\includegraphics[scale=0.84]{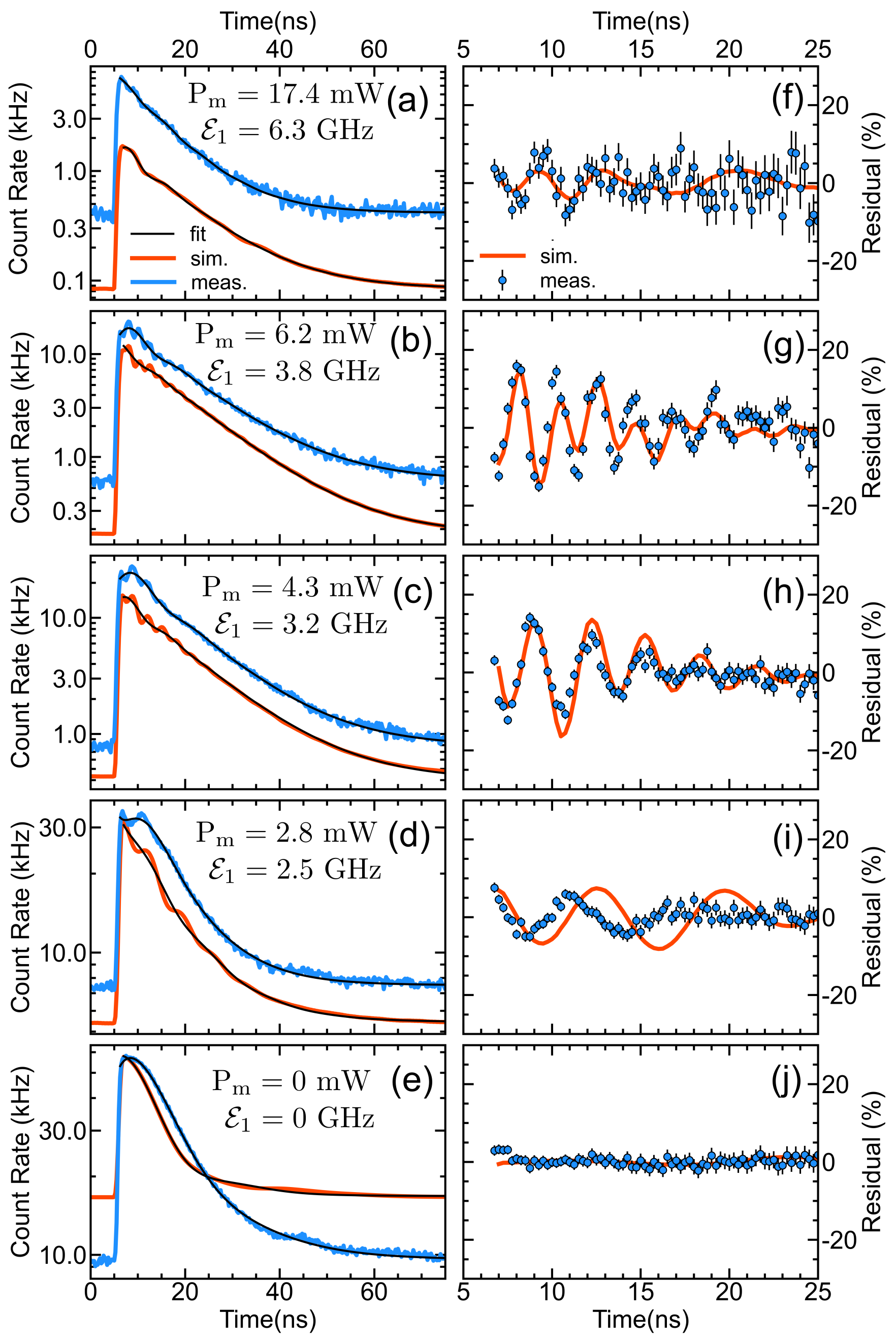}
\caption{\textbf{Comparing simulated time-domain to experiment.} \textbf{(a-e)} Histogram of time-tagged photon counts relative to a repeated 1 ns duration red laser pulse at $\Delta$ = 0 for several mechanical drive powers (blue trace) and simulated photon count rates for comparable $\mathcal{E}_1$ drive strength. Fits (black traces) are to Equation \ref{equation: sup5}. \textbf{(f-j)} Extracted residual oscillations for each experimental (blue points) and simulated (red trace) curve in \textbf{a-e}. Error bars in (f-j) are determined using the shot-noise limit.}\label{fig:Figure_S4_version_1}
\end{figure}

With all simulation parameters for the Hamiltonian in Equation \ref{equation: sup2} now determined, we simulate the orbital Rabi measurements. We apply the same spectral diffusion model as for the optical Rabi simulation. We apply two 1 ns duration laser pulses separated by 100 ns for various mechanical drive amplitudes and simulate the density matrix out to 200 ns. In Figure \ref{fig:Figure_S4_version_1} we plot the simulated time-domain PL of the second laser pulse against the measured time-domain data. Here we use a contrast of $\beta$ = 0.6 and apply a global scaling to the simulation data such that the maximum simulated PL with no mechanical drive matches the maximum measured PL for no drive (Figure~\ref{fig:Figure_S4_version_1} panel (e)).

Our time domain simulations show good qualitative agreement with the measurement. In particular, when no mechanical drive is applied we find qualitatively similar PL response in measurement and simulation. The rounded excited state decay is a direct result of the imperfect laser extinction and spectral diffusion. The imperfect extinction causes the system to undergo slow Rabi oscillations even when the EOM is gated off, and the effect of spectral diffusion is to provide an inhomogeneous rate for such oscillations which results in a damping. The black curves in Figure \ref{fig:Figure_S4_version_1}(a-e) are fits to:

\begin{equation}\label{equation: sup5}
y(t) = a + b e^{-t / \tau_f} + c \cos (\omega_{k} t + \phi) e^{-t/ \tau_k}
\end{equation}
\
where $a$ gives the overall steady-state background PL, $b e^{-t / \tau_f}$ describes the expected spontaneous emission from the excited state, and $c \cos (\omega_{k} t + \phi) e^{-t/ \tau_k}$ captures the slow oscillation which results from spectral diffusion and imperfect laser extinction. Given that this model fits well to the optical response when no mechanical drive is applied to the system, we use this model to extract the residual oscillations present in the PL response when the system is mechanically driven. To avoid capturing the mechanical oscillations with the third term in Equation~\ref{equation: sup5}, we constrain $\omega_k$ to be at most 100 MHz and $\tau_k$ to be at least 5 ns.

Extracted residual oscillations for the measurement and simulation are shown in Figure~\ref{fig:Figure_S4_version_1}(f-j). The drive amplitude scaling of both the frequency of oscillations and the visibility of the oscillations are in good agreement. We conclude from this that the model used to extract the residuals accurately accounts for the EOM and spectral diffusion effects, while preserving the orbital oscillations and their decay.

\begin{figure}
\centering
\includegraphics[scale=0.84]{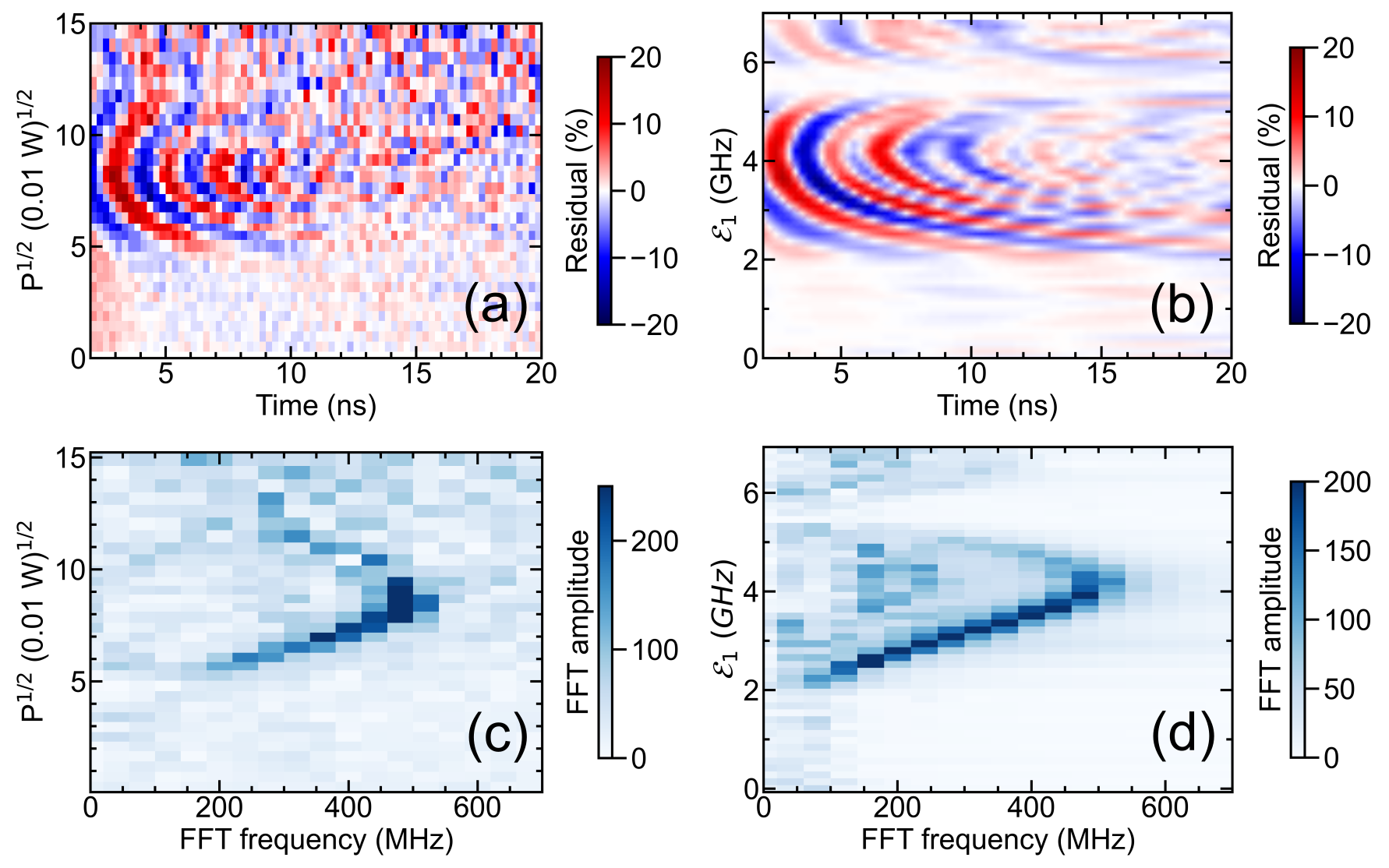}
\caption{\textbf{FFT of measured and simulated residuals.} \textbf{(a)} Measured and \textbf{(b)} simulated residual traces versus mechanical drive power. \textbf{(c)} FFT of measured and \textbf{(d)} simulated residual traces showing qualitatively similar oscillation frequency versus power.}\label{fig:Figure_S5_version_1}
\end{figure}

Additionally, we examine the frequency of orbital oscillations by taking the FFT of the measured and simulated residuals in Figure \ref{fig:Figure_S5_version_1}. There is good agreement between the dominant FFT peak of the simulated and measured residual oscillations indicating that our simulations capture the same physics as the measurement.

\section{Orbital Rabi decoherence simulations}\label{section: Orbital Rabi decoherence simulations}

We model the orbital Rabi decoherence process via simulation. We first generate a random set of 500 Gaussian-distributed electric fields along the $\hat{x}$ and $\hat{y}$ directions. The distributions each have standard deviations of 35 MHz. For each $(\hat{x},\hat{y})$ pair of electric fields we calculate the time-domain response of the system oscillating between $\ket{E_{\textrm{x}}}$ and $\ket{E_{\textrm{y}}}$, assuming that $P_y=\Omega_0^2/(\Omega_0^2+\delta^2) (1-\cos{\Omega_R t})/2$, where $\Omega_0$ and $\delta$ are given by Equations 4 and 5 in the main text. We calculate the average time-domain response of the system across the full set of fluctuations and fit a decaying sinusoid to characterize the decoherence time. The results are shown as a family of solid curves in Figure~4(b) of the main text. As mentioned there, the predicted decoherence versus drive power for the measured defect strain splitting of $\Delta = 6.41$ GHz (Figure~4(b) of the main text, black trace) is a poor fit to the experiment, but a +1.5\% change in $\Delta$ (Figure~4(b) of the main text, red trace) produces a similar trend to the measurement. The primary source of error in our characterization of $\Delta$ is the Fabry-Perot cavity which is used to monitor the resonant laser frequency shift as we tune the laser in spectroscopy. The error in such a measurement is approximately 2-3\% given our cavity's FSR (10 GHz), finesse ($\sim$150), and our procedure for converting the laser's piezo tuning voltage to frequency detuning.

\section{Modeling $\Omega_{\textrm{R}}$}

As demonstrated in our experiment, 
dynamic strain 
provides
significant control over the orbital degrees of freedom of 
a nitrogen-vacancy (NV) center.  For example, it allows us  to coherently move population between the two excited states $|E_x\rangle,|E_y\rangle$.  Here we model this process.  In Sec.~\ref{pe} we use perturbation theory, within a Floquet picture, to calculate the properties of the excited state manifold, including the Rabi splitting $\Omega_R$.  In Sec.~\ref{spec} we review the relation between the optical absorption spectrum and the Floquet eigenstates calculated in Sec.~\ref{pe}. In Sec.~\ref{subsection: Landau-Zener Transfer Matrix} we give a strong-coupling expansion based upon repeated Landau-Zener sweeps.


In the basis $\{ |{E_{\textrm{x}}\rangle}, 
|{E_{\textrm{y}}}\rangle \}$, the Hamiltonian describing the interaction of the NV center with the static strain of the diamond and the dynamic strain driven by our resonator is given by:
\begin{eqnarray}\label{equation: sup6}
\hat{H}(t) &=& H_{A_1}(t) \hat{\mathbbm{1}} + H_{E_1}(t) \hat{\sigma}_{z} + H_{E_2} \hat{\sigma}_{x} \\\nonumber
&=& \left(
\begin{array}{cc}
H_{A_1}(t)+H_{E_1}(t)&H_{E_2}(t)\\
H_{E_2}(t)&H_{A_1}(t)-H_{E_1}(t)
\end{array}\right)
\end{eqnarray}
where $\hat{\mathbbm{1}}$, $\hat{\sigma}_{z}$, and $\hat{\sigma}_{x}$ are the identity matrix,  and  Pauli matrices. The Hamiltonian components $H_{i}$ of Jahn-Teller symmetry $i$ have both static and dynamic components, given by:
\begin{align}\label{equation: sup7}
H_{A_1}(t) &=  V_{A_1} + \mathcal{A}_1 \cos (\omega_{\textrm{m}} t ) \\ 
H_{E_1}(t) &=  V_{E_1} + \mathcal{E}_1 \cos (\omega_{\textrm{m}} t ) \\ 
H_{E_2} &= V_{E_2}.
\end{align}
Here $V_{i}$ are static strain potentials resulting from the local strain environment of the defect, $\mathcal{A}_1$ and $\mathcal{E}_1$ are the dynamic strain amplitudes driven by the HBAR, and $\omega_{\mathrm{m}}$ is the mechanical drive frequency.  
The $V_{i}$, $\mathcal{A}_1$, and $\mathcal{E}_1$ are different for every NV center, and need to be determined experimentally.


By Floquet's theorem \cite{floquet}, the generic solution to Schrodinger's equation $i\partial_t |\psi(t)\rangle=H|\psi(t)\rangle$ is 
\begin{equation}\label{fexp}
|\psi(t)\rangle = a |\phi_1 (t)\rangle + b |\phi_2(t)\rangle
\end{equation}
where $|\phi_j(t)\rangle$ are Floquet eigenstates, obeying $|\phi_j(t+T)\rangle=e^{-i \nu_j T} |\phi_j(t)\rangle$. Here $T=2\pi/\omega_m$ is the periodicity of the drive and $\nu_j$ are the Floquet quasi-energies.  As detailed in Sec.~\ref{spec}, this quasiperidoc structure leads to optical absorption at frequencies $\nu_j+n\omega_m$, for all integers $n$.  The Floquet quasi-energies are only defined up to a multiple of $\omega_m$, and can be chosen so that  $|\nu_j|<\omega_m/2$.  This is the spectral analogy of working in the first Brillioun zone.  The Rabi splitting is then defined as $\Omega_R=(\nu_1-\nu_2)$.

In the absence of the drive, the Hamiltonian in Eq.~(\ref{equation: sup6}) is static with eigenvalues $\epsilon_{\pm}=V_{A_1}\pm \sqrt{V_{E_1}^2+V_{E_2}^2}$.  We consider the near-resonant limit where ${\Delta=}\epsilon_+-\epsilon_-\approx n \omega_m$ for some integer $n$.  In our experiment $n=5$.  To match our Floquet definitions, we shift these energies by multiples of $\omega_m$.  The undriven Rabi splitting is then $\Omega_R^0=(\epsilon_+-\epsilon_--n\omega_m)=2 \sqrt{V_{E_1}^2+V_{E_2}^2}-n\omega_m$.


\subsection{Perturbative expansion}\label{pe}

The particular NV center in our experiment has $V_{E_2}\ll V_{E_1}$ which allows us to derive simple expressions for $\Omega_R$.  We begin by transforming into a rotating frame, writing
\begin{equation}\label{rot}
|\psi\rangle = \left(\begin{array}{c}u(t) e^{-i(\Phi(t)+\Theta(t))}\\
v(t) e^{-i(\Phi(t)-\Theta(t))}
\end{array}\right),
\end{equation}
with 
$\partial_t\Phi=H_{A_1}$ and 
$\partial_t\Theta=n\omega_m/2+\mathcal{E}_1  \cos (\omega_m t )$, corresponding to
\begin{eqnarray}
\Phi(t)&=&V_{A_1} t + \frac{{\cal A}_1}{\omega_m}\sin \omega_m t\\
\Theta(t)&=&\frac{n\omega t}{2}+\frac{\mathcal{E}_1}{\omega}  \sin\omega_m t.
\end{eqnarray}
Substituting this ansatz into the Schrodinger equation, $i\partial_t |\psi\rangle=\hat H|\psi\rangle$, leads to
\begin{equation}\label{time}
i\partial_t \left(\begin{array}{c} u(t)\\v(t)\end{array}\right)=
\left(\begin{array}{cc} \delta_0&
\Omega(t)\\
\Omega^*(t)&-\delta_0\end{array}\right)\left(\begin{array}{c} u(t)\\v(t)\end{array}\right).
\end{equation}
The transformation was chosen so that the detuning $\delta_0=V_{E_1}-n\omega_m/2$ is small, and so that the transition matrix element is periodic,
\begin{equation}
\Omega(t)=V_{E_2} \exp i\left(
n\omega_m t + \frac{2{\cal E}_1}{\omega_m}\sin(\omega_m t)
\right).
\end{equation}
Note that $|u(t)|^2$ and $|v(t)|^2$ are the probabilities of being in the states $|E_x\rangle$ and $|E_y\rangle$.  Using the Jacobi-Anger expansion, we can write
the Fourier expansion $
\Omega(t)=\sum_j e^{ij\omega_m t} 
\Omega_j$ 
with
\begin{equation}
\Omega_j = V_{E_2} J_{j-n}\left(\frac{2\mathcal{E}_1}{\omega_{\textrm{m}}} \right)
\end{equation}
where $J_j(x)$ is the Bessel function of order $j$.

We then make the Floquet ansatz, $u(t)=\sum_j e^{-i(j\omega_m+\nu) t} u_j$ and $v(t)=\sum_j e^{-i(j\omega_m+\nu) t} v_j$, where
 $\nu$ is the  quasi-energy, introduced after Eq.~(\ref{fexp}).
Substituting this ansatz into Eq.~(\ref{time}), yields
\begin{eqnarray}\label{uv}
\nu u_j&=& (\delta_0-j\omega_m) u_j+\sum_s 
\Omega_s v_{j+s}\\
\nu v_j&=& (-\delta_0-j\omega_m) v_j+\sum_s 
\Omega_{s} u_{j-s}.
\end{eqnarray}
These equations are exact, as we have not yet made any approximations.  The Floquet eigenstates can numerically be found by truncating to a finite set of $\{j\}$ and solving the resulting linear algebra problem.   One can understand $\Omega_s$ as the transition matrix element coming from absorbing $n+s$ quanta from the resonator.  Due to the resonance condition, the leading contribution comes from $s=0$.

Under the assumption that $\delta$ and $\Omega$ are small, we can analytically solve Eq.~(\ref{uv}). The key insight is that we can find solutions where $u_{j\neq 0}$ and $v_{j\neq 0}$ are small.  To lowest order we set $u_{j\neq 0}=0$ and $v_{j\neq 0}=0$ to arrive at
\begin{equation}
\nu\left(
\begin{array}{c}
u_0\\
v_0
\end{array}
\right)= \left(
\begin{array}{cc}
\delta_0&
\Omega_0\\
\Omega_0&-\delta_0
\end{array}
\right)
\left(
\begin{array}{c}
u_0\\
v_0
\end{array}
\right)\qquad \mbox{(first order)}
\end{equation}
One readily sees that to this order 
$\nu=\pm\sqrt{\delta_0^2+
\Omega_0^2}$.  

To derive the second order result, we consider Eq.~(\ref{uv}) for $j\neq 0$, and on the right hand side neglect terms $u_{i}$ or $v_{i}$, where $i\neq0,j$. This gives $u_j= v_0 \Omega_{-j}/(\nu+j\omega_m-\delta)$ and $v_j=u_0 \Omega_{j}/(\nu+j\omega_m+\delta)$.  Since both $\nu$ and $\delta$ are small these further simplify to $u_j=v_0 \Omega_{-j}/(j\omega_m)$ and $v_j=u_0 \Omega_{j}/(j\omega_m)$.  Substituting these back into the $j=0$ equations gives
\begin{equation}\label{secorder}
\nu \left(\begin{array}{c}u_0\\v_0\end{array}\right)
= \left(
\begin{array}{cc}
\delta&
\Omega_0\\
\Omega_0&-\delta
\end{array}
\right)
\left(\begin{array}{c}u_0\\v_0\end{array}\right)
\qquad\mbox{(second order)}
\end{equation}
where
\begin{equation}
\delta=V_{E_1}-\frac{n\omega_m}{2}+\sum_{s\neq 0} 
\frac{\Omega_{s}^2}{s\omega_m}
\end{equation}
The quasi-energies are then $\nu=\pm \sqrt{
\Omega_0^2+ \delta^2}$ and the Rabi frequency is  $\Omega_R=(\nu_+-\nu_-)=2\sqrt{
\Omega_0^2+ \delta^2}$.  
Similar expansions can be found in references
\cite{RodriguezVega2018,Mananga2011}.

Since $u_{j\neq0}$ and $v_{j\neq 0}$ are small, we can interpret $|u_0|^2$ and $|v_0|^2$ as the probabilities of occupying the states $|E_x\rangle$ and $|E_y\rangle$.   At the same level of approximation, we can calculate dynamics in the $E_{xy}$ manifold by substituting $\nu\to i\partial_t$ in Eq.~(\ref{secorder}).

For small ${\cal E}_1$, $\Omega_s\to V_{E_2}({\cal E}_1/{\omega_m})^{(s-n)}/
\Gamma(s-n+1) $, which vanishes except for $s=n$.  Thus in this limit the sum in $\delta$ consists of a single term and we have $\delta\to\delta_0+ V_2^2/(n\omega)$.  For large $\cal E$ we find
\begin{eqnarray}
\delta({\cal E}_1\to\infty)&\to&\delta_0+ \frac{V_{E_2}^2 \omega_m}{{\cal E}_1^2}
\sin\left(
4\frac{\mathcal{E}_1}{\omega_m}+\frac{n^2\omega_m}{2\mathcal{E}_1}+\left(n+\frac{1}{2}\right)\frac{\pi}{2}
\right)\\
\Omega_0(\mathcal{E}_1\to\infty)&\to& \frac{V_{E_2}}{\sqrt{\pi}}\sqrt{\frac{\omega_m}{\mathcal{E}_1}}\cos\left(
2 \frac{\mathcal{E}_1}{\omega_m}+\frac{n^2\omega_m}{4\mathcal{E}_1}+\left(n-\frac{1}{2}\right)\frac{\pi}{2}
\right),
\end{eqnarray}
the latter of which can be derived from standard Bessel function identities.


\subsection{Spectroscopy}\label{spec}

We now relate the solutions of the two-level problem in Sec.~\ref{pe} to the absorption rate of a laser which is near resonant for a transition between the NV ground state $|0\rangle$ and the excited states $|E_x\rangle$ and $|E_y\rangle$.  In a rotating wave approximation, this optical field can be modelled as $
\hat H_d=\hat A_+ e^{i\omega_d t} +\hat A_- e^{-i\omega_d t}$, where $\omega_d$ is the photon frequency and $\hat A_+=(\hat A_-)^\dagger=\Omega_x |E_x\rangle\langle 0| + \Omega_y |E_y\rangle\langle 0|$.  The Rabi frequencies $\Omega_x$ and $\Omega_y$ are related to the laser amplitude and polarization.

The operator corresponding to the net transition rate is
$\hat J=\partial_t |0\rangle \langle 0|=-i[|0\rangle \langle 0|,\hat H_d]$, which evaluates to
$
\hat J(t) = i e^{i\nu t} \hat A_+ -i e^{-i\nu t}\hat A_-.$
The transition rate from the ground state can then formally be expressed as
\begin{equation}
\langle \hat J(t)\rangle =
\langle 0 | T\!\exp\left(i\int_{t_0}^{t}[ \hat H(\tau)+\hat H_d(\tau)]d\tau\right) \hat J(t) \,\,T\!\exp\left(-i\int_{t_0}^{t} [H(\tau)+\hat H_d(\tau)]d\tau\right)|0\rangle
\end{equation}
where ``$T\!\exp$'' denotes the time ordered exponential and  $\hat H(\tau)$ is the Hamiltonian for the excited state manifold, given by Eq.~(\ref{equation: sup6}).  Here $t_0$ is the time at which the drive is turned on: We will take  $t_0\to-\infty$.

Under the assumption that $\Omega_{\alpha}$ is small we can expand the time evolution operators to leading order in $\hat H_d$, resulting in
\begin{eqnarray}
\langle \hat J(t)\rangle&=&
\int_{t_0}^t\!\!d\bar t\,\left[
\langle 0 |
Te^{i\int_{\bar t}^{t}\hat H(\tau)d\tau}
(i H_d(\bar t))\,
Te^{i\int_{t_0}^{\bar t}\hat H(\tau)d\tau}
\hat J(t)
\,Te^{-i\int_{t_0}^{t}\hat H(\tau)d\tau}|0\rangle
\right.
\\\nonumber &&\qquad
\left.+
\langle 0 |
Te^{-i\int_{t_0}^{t}\hat H(\tau)d\tau}
\hat J(t)
Te^{-i\int_{\bar t}^{t}\hat H(\tau)d\tau}
(-iH_d(\bar t))
Te^{-i\int_{t_0}^{\bar t}\hat H(\tau)d\tau}
|0\rangle
\right]
\end{eqnarray}
Noting that $\hat H$ acts trivially on $|0\rangle$, and using the explicit expressions for $\hat H_d$ and $\hat J$ we can simplify this to
\begin{eqnarray}
\langle \hat J(t)\rangle
\label{jeq}
&=&2 \sum_{\alpha\beta}\Omega_\alpha^*\Omega_\beta\,{\rm Im} \int_{-\infty}^\infty\!d\tau\,e^{-i\omega_d (t-\tau)} C_{\alpha\beta}(t,\tau)
\end{eqnarray}
where $\alpha$ and $\beta$ are summed over $\{X,Y\}$ and
\begin{equation}\label{cf}
C_{\alpha\beta}(t_1,t_2)
=\frac{1}{i}\theta(t_1-t_2)\langle E_\alpha|\, T\exp\left({-i\int_{t_1}^{t_2} \hat H(\tau)d\tau}\right)|E_\beta\rangle
\end{equation}
 We can calculate the correlation function in Eq.~(\ref{cf}) in terms of the Floquet eigenstates $|\phi_j(t)\rangle$, which form an orthonormal basis and have the property
\begin{equation}
T\exp\left({-i\int_{t_1}^{t_2} \hat H(\tau)d\tau}\right)
|\phi(t_1)\rangle=|\phi(t_2)\rangle
\end{equation}
This allows us to insert a resolution of the identity to conclude
\begin{equation}\label{cpsi}
C_{\alpha\beta}(t_1,t_2)=\frac{1}{i}\theta(t_1-t_2)\sum_j \langle E_\alpha|\phi_j(t_2)\rangle \langle \phi_j(t_1)|E_\beta\rangle.
\end{equation}
In Sec.~\ref{pe} we calculated the spectral expansion
\begin{equation}
|\phi_j(t)=e^{-i\epsilon_j t}\sum_n e^{-i n \omega_m t}\left(\phi_{nj}^X |E_X\rangle+ \phi_{nj}^Y |E_Y\rangle\right).
\end{equation}
We substitute this ansatz into Eq.~(\ref{cpsi}), and perform the integrals in Eq.~(\ref{jeq}) to find
\begin{equation}
\langle \hat J\rangle 
= -\sum_{\alpha,\beta} \Omega_\alpha^*\Omega_\beta \sum_{jn} (\phi^\alpha_{nj})^* \phi^\beta_{nj}\, 2\pi \delta(\epsilon_j+n\omega_m-\omega_d) +\cdots
\end{equation}
where the neglected terms rapidly oscillate with zero mean.  Thus we have spectral weight when $\omega_d=\epsilon_j+n\omega_m$ for integer $n$.  The coefficients $\phi_{nj}^\alpha$ are related to the $u_j$ and $v_j$ in Sec.~\ref{pe} by the transformation in Eq.~(\ref{rot})

\subsection{Landau-Zener Transfer Matrix}\label{subsection: Landau-Zener Transfer Matrix}

In the regime of strong driving where $\mathcal{E}_1 \gg V_{E_2}$, we can use a transfer matrix approach to reliably calculate the Rabi oscillation rate. The transfer matrix approach assumes that the transitions between the orbital states are of the form of Landau-Zener tunnelings.

We can intuitively understand the transfer matrix approach by considering one cycle of the mechanical drive. Closer to the antinodes of the cosine drive, when $|\mathcal{E}_1 \cos{\omega_m t}| \gg V_{E_2}$, there are no transitions between the orbital states and the states only pick up dynamical phases. Near the nodes of the drive where $|\mathcal{E}_1 \cos{\omega_m t}| \ll V_{E_2}$, the orbital states are nearly degenerate and the drive sweeps through it with a speed $\mathcal{E}_1 \omega_m$. There can be Landau-Zener transitions during this sweep at a rate given by $\sin^2 \chi/2 = 1- \exp{(-4\pi V_{E_2}^2/(2\mathcal{E}_1 \omega_m) )}$. 

We use this transfer matrix approach as outlined in~\cite{Ashhab_2007} to calculate the Rabi oscillation rate in the high driving limit. We find that the Rabi frequency in terms of our parameters is given by, 
\begin{equation}\label{equation: sup13}
    \Omega_R = \omega_m \sin \frac{\chi}{2}|\cos (\theta - \theta_{stokes})|
\end{equation}
In the equation above, 
\begin{equation}\label{equation: sup14}
    \theta = \frac{2\sqrt{\mathcal{E}_1^2- V_{E_1}^2}}{\omega_m} - \frac{2V_{E_1}}{\omega_m}\cos^{-1}{\frac{V_{E_1}}{\mathcal{E}_1}}
\end{equation}
\begin{equation}\label{equation: sup15}
    \theta_{stokes} = \frac{\pi}{4} + \Gamma(1-i\eta) + \eta(\log \eta - 1)
\end{equation}
Here $\eta = V_{E_2}^2/(2\mathcal{E}_1\omega_m)$ and $\Gamma(x)$ is the gamma function.

\section{Extended Time-Domain Data and Fitting}\label{section: Extended Time-Domain Data and Fitting}

\begin{figure*}%
\centering
\includegraphics[scale=0.70]{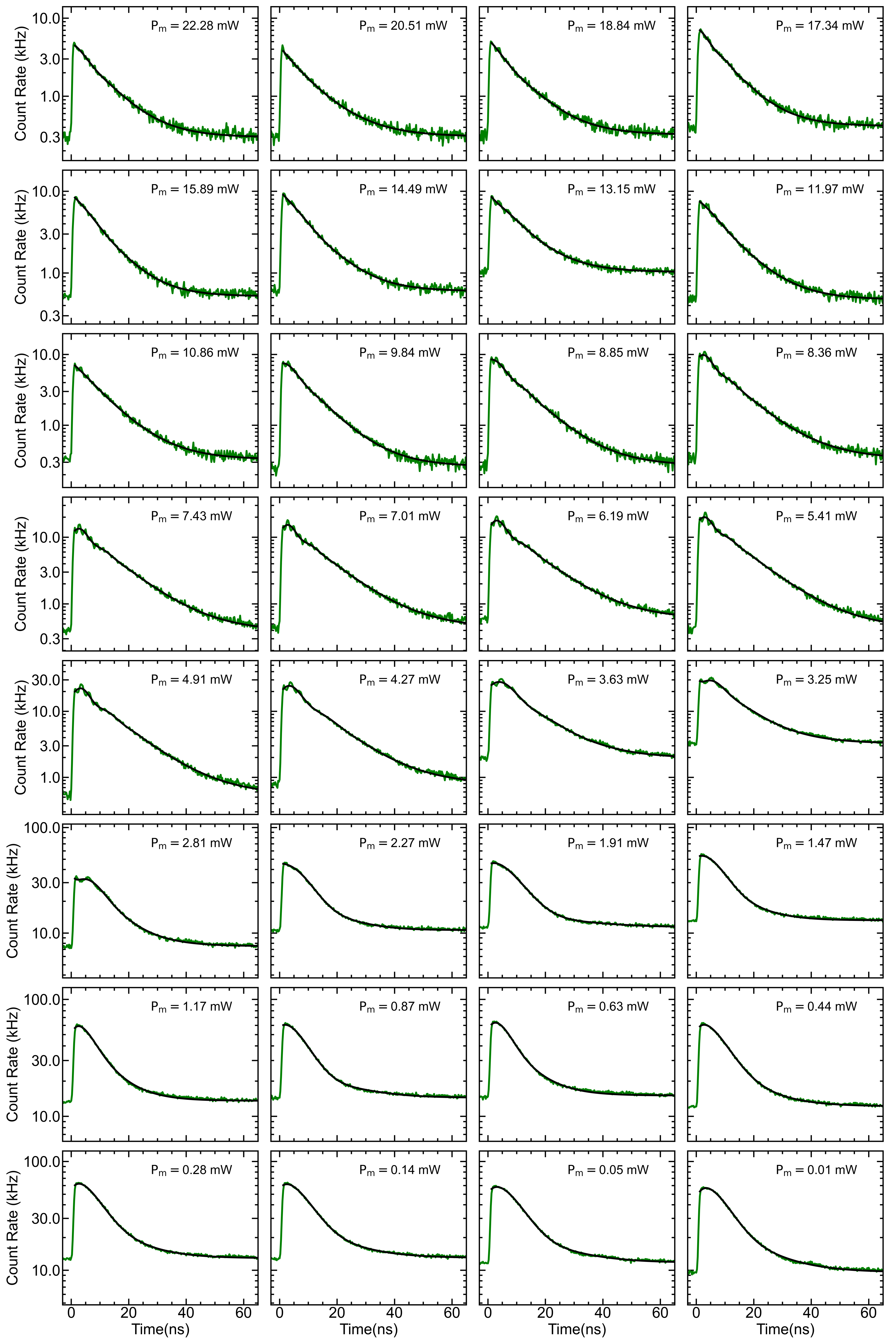}
\caption{\textbf{Extended experimental time-domain traces.} Histogram of time-tagged photon counts relative to a repeated 1 ns duration red laser pulse at $\Delta$ = 0 for all measured mechanical drive powers (green traces). Fits (black traces) are to Equation \ref{equation: sup5}.}\label{fig:Figure_S6_version_1}
\end{figure*}

For completeness we include the measured time-resolved data and fits to Equation~\ref{equation: sup5} for all mechanical drive powers in Figure~\ref{fig:Figure_S6_version_1}. By inspection the fits used to extract the residual orbital oscillations are reasonable for all powers.

\begin{figure*}%
\centering
\includegraphics[scale=0.70]{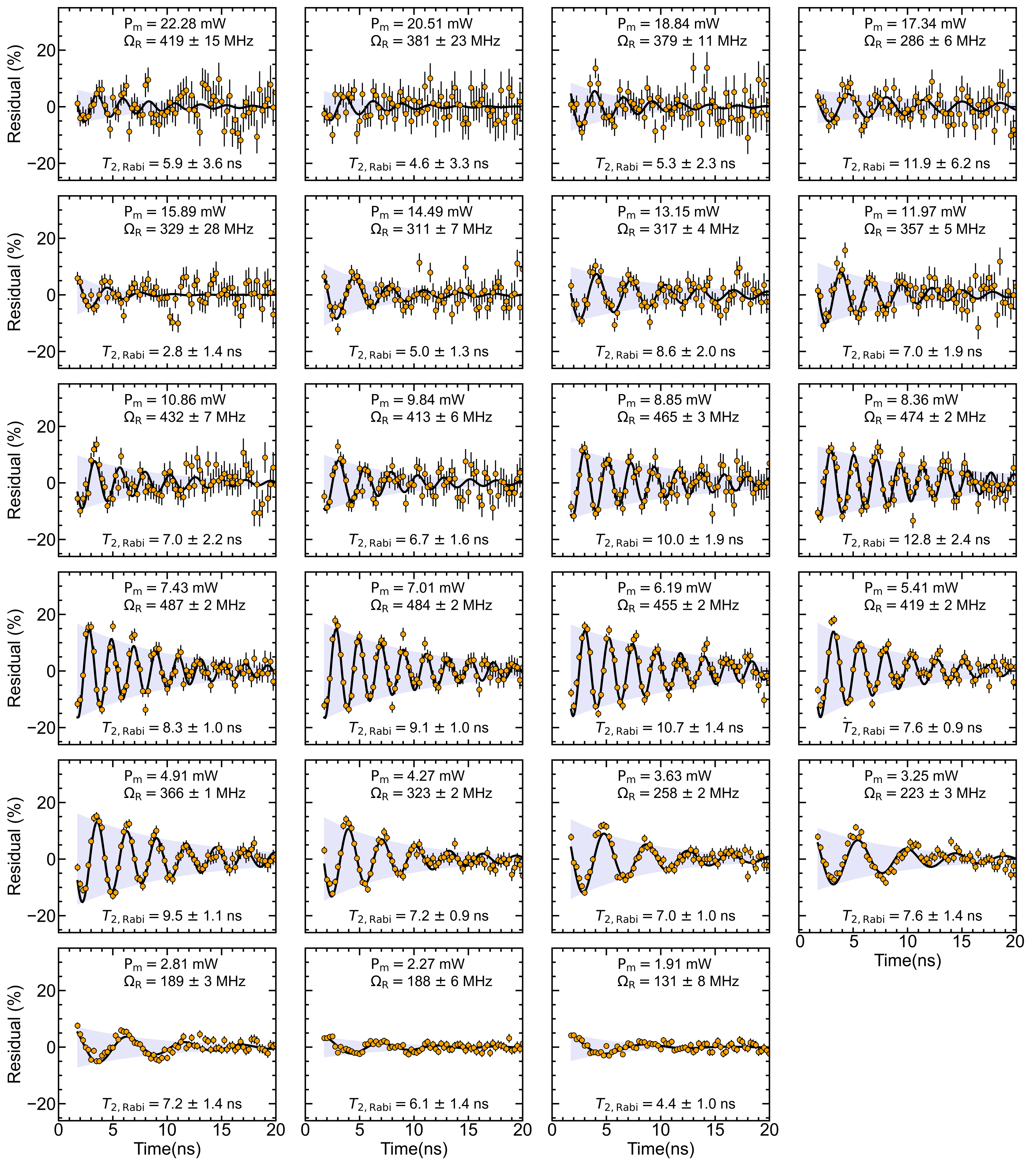}
\caption{\textbf{Extended experimental time-domain residuals.} Extracted residual oscillations for all measured powers where an obvious orbital oscillation is present (orange points). Fits (black) traces are to $y(t) = A \cos (\Omega_{\textrm{R}} t) e^{-t / T_{2,\textrm{Rabi}}}$. Error bars on individual points are determined using the shot-noise limit.}\label{fig:Figure_S7_version_1}
\end{figure*}

We additionally include the residual orbital oscillations and fits to a decaying sine response for all mechanical drive powers in~\ref{fig:Figure_S7_version_1}.

\clearpage

\bibliography{bibliography}